%% file: main.tex
\documentclass{ieeeaccess}

\usepackage{graphicx}
\usepackage{amsmath,amssymb,amsfonts}
\usepackage{textcomp}
\usepackage{bm}
\usepackage{physics}
\usepackage{algorithm}
\usepackage{algpseudocode}
\usepackage{array}
\usepackage{slashbox}
\usepackage{dcolumn}
\usepackage{siunitx}
\usepackage{cite}
\usepackage{booktabs}
\usepackage{framed} 

\makeatletter
\let\MYcaption\@makecaption
\makeatother

\usepackage{subcaption}

\makeatletter
\let\@makecaption\MYcaption
\makeatother

\makeatletter
\AtBeginDocument{\DeclareMathVersion{bold}
\SetSymbolFont{operators}{bold}{T1}{times}{b}{n}
\SetSymbolFont{NewLetters}{bold}{T1}{times}{b}{it}
\SetMathAlphabet{\mathrm}{bold}{T1}{times}{b}{n}
\SetMathAlphabet{\mathit}{bold}{T1}{times}{b}{it}
\SetMathAlphabet{\mathbf}{bold}{T1}{times}{b}{n}
\SetMathAlphabet{\mathtt}{bold}{OT1}{pcr}{b}{n}
\SetSymbolFont{symbols}{bold}{OMS}{cmsy}{b}{n}
\renewcommand\boldmath{\@nomath\boldmath\mathversion{bold}}}
\DeclareFontShape{T1}{formata}{m}{sl}{<-> ssub * formata/m/it}{}
\DeclareFontShape{T1}{formata}{b}{sl}{<-> ssub * formata/b/it}{}
\makeatother

\def\BibTeX{{\rm B\kern-.05em{\sc i\kern-.025em b}\kern-.08em
    T\kern-.1667em\lower.7ex\hbox{E}\kern-.125emX}}

\begin{document}
\history{Date of publication xxxx 00, 0000, date of current version xxxx 00, 0000.}
\doi{10.1109/ACCESS.2024.0429000}

\onecolumn
\begin{framed}
    \noindent
    This work has been submitted to the IEEE for possible publication. Copyright may be transferred without notice, after which this version may no longer be accessible.%
\end{framed}
\clearpage

\title{Extending Sample Persistence Variable Reduction for Constrained Combinatorial Optimization Problems}
\author{\uppercase{Shunta Ide}\authorrefmark{1},
    \uppercase{Shuta Kikuchi}\authorrefmark{2, 3}, and Shu Tanaka\authorrefmark{1,2,3,4},\IEEEmembership{Member, IEEE}
}

\address[1]{Department of Applied Physics and Physico-Informatics, Keio University, Yokohama, Kanagawa 223-8522, Japan}
\address[2]{Graduate School of Science and Technology, Keio University, Yokohama, Kanagawa 223-8522, Japan}
\address[3]{Keio University Sustainable Quantum Artificial Intelligence Center (KSQAIC), Keio University, Minato, Tokyo 108-8345, Japan}
\address[4]{Human Biology-Microbiome-Quantum Research Center (WPI-Bio2Q), Keio University, Shinjuku, Tokyo 160-8582, Japan}
\tfootnote{This work was partially supported by the Japan Society for the Promotion of Science (JSPS) KAKENHI (Grant Number JP23H05447), the Council for Science, Technology, and Innovation (CSTI) through the Cross-ministerial Strategic Innovation Promotion Program (SIP), ``Promoting the application of advanced quantum technology platforms to social issues'' (Funding agency: QST), Japan Science and Technology Agency (JST) (Grant Number JPMJPF2221).}

\markboth
{Ide \headeretal: Extending Sample Persistence Variable Reduction for Constrained Combinatorial Optimization Problems}
{Ide \headeretal: Extending Sample Persistence Variable Reduction for Constrained Combinatorial Optimization Problems}

\corresp{Corresponding author: Shunta Ide (e-mail: shunta16213@keio.jp).}

\begin{abstract}
    Constrained combinatorial optimization problems (CCOPs) are challenging to solve due to the exponential growth of the solution space.
    When tackled with Ising machines, constraints are typically enforced by the penalty function method, whose coefficients must be carefully tuned to balance feasibility and objective quality.
    Variable-reduction techniques such as sample persistence variable reduction (SPVAR) can mitigate hardware limitations of Ising machines, yet their behavior on CCOPs remains insufficiently understood.
    Building on our prior proposal, we extend and comprehensively evaluate multi-penalty SPVAR (MP-SPVAR), which fixes variables using solution persistence aggregated across multiple penalty coefficients.
    Experiments on benchmark problems, including the quadratic assignment problem and the quadratic knapsack problem, demonstrate that MP-SPVAR attains higher feasible-solution ratios while matching or improving approximation ratios relative to the conventional SPVAR algorithm.
    An examination of low-energy states under small penalties clarifies when feasibility degrades and how encoding choices affect the trade-off between solution quality and feasibility.
    These results position MP-SPVAR as a practical variable-reduction strategy for CCOPs and lay a foundation for systematic penalty tuning, broader problem classes, and integration with quantum-inspired optimization hardware, as well as quantum algorithms.
\end{abstract}

\begin{keywords}
    Combinatorial optimization, constrained combinatorial optimization, Ising machine, penalty function, quadratic assignment problem, quadratic knapsack problem, sample persistence, variable reduction
\end{keywords}

\titlepgskip=-21pt

\maketitle

\section{Introduction}\label{sec:introduction}
\input{sections/introduction/introduction.tex}

\section{Method}\label{sec:method}
\input{sections/method/method.tex}

\section{Experimental Setup}\label{sec:experimental-setup}
\input{sections/experimental-setup/experimental-setup.tex}

\section{Results}\label{sec:result}
\input{sections/result/result.tex}

\section{Discussion}\label{sec:discussion}
\input{sections/discussion/discussion.tex}

\section{Conclusion and Future Work}\label{sec:conclusion}
\input{sections/conclusion/conclusion.tex}

\appendices
\input{sections/appendicies/appendicies.tex}

\section*{Acknowledgments}
S. Tanaka wishes to express their gratitude to the World Premier International Research Center Initiative (WPI), MEXT, Japan, for their support of the Human Biology-Microbiome-Quantum Research Center (Bio2Q).

\bibliographystyle{IEEEtran}
\bibliography{IEEEabrv,main}

\begin{IEEEbiography}[{\includegraphics[width=1in,height=1.25in,clip,keepaspectratio]{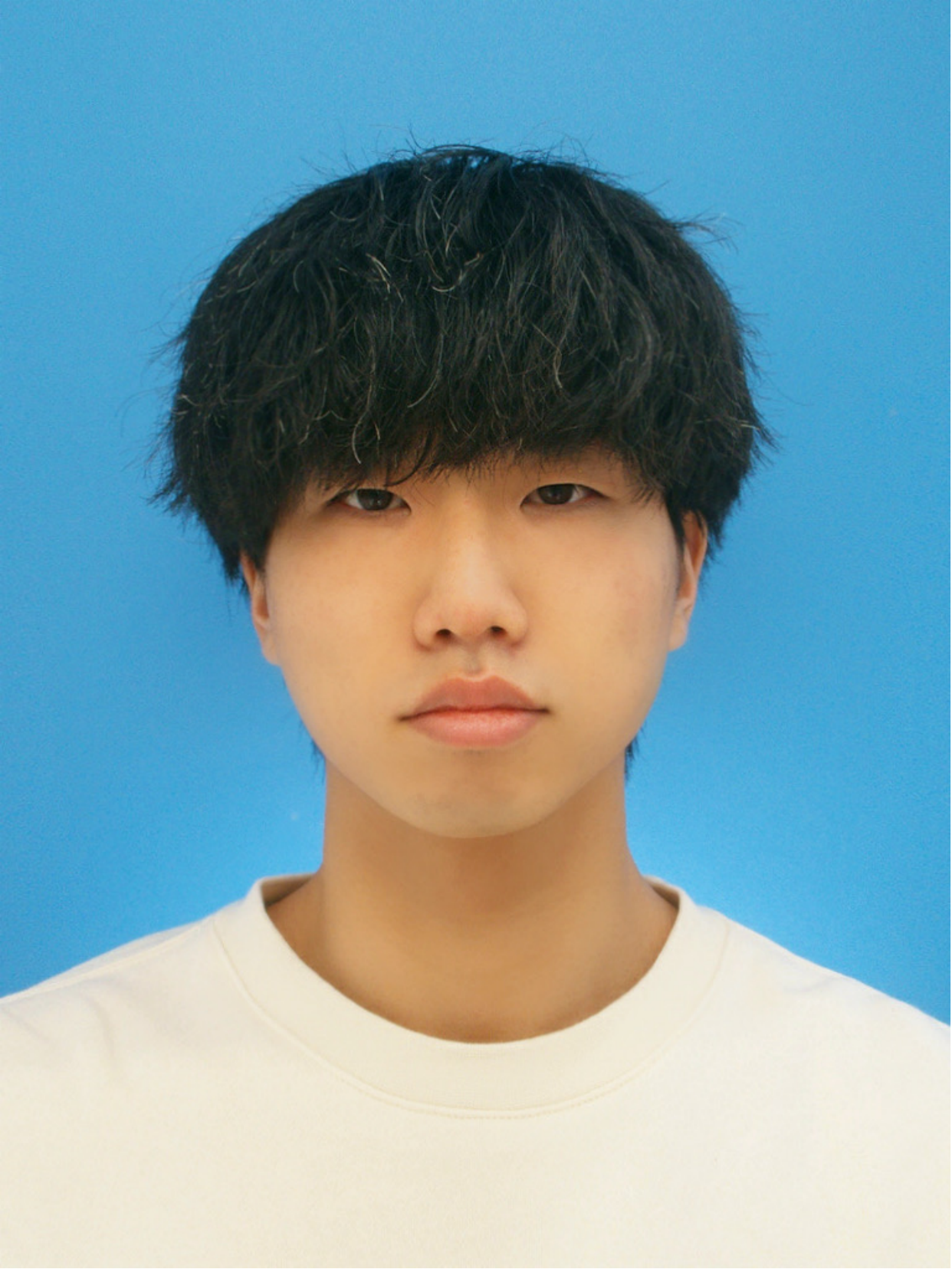}}]{Shunta Ide}
    is currently pursuing the B.~Eng. degree in the Department of Applied Physics and Physico-Informatics at Keio University, Kanagawa, Japan.
    His research interests include Ising machines, quantum annealing, quantum computing, and combinatorial optimization.
\end{IEEEbiography}

\begin{IEEEbiography}[{\includegraphics[width=1in,height=1.25in,clip,keepaspectratio]{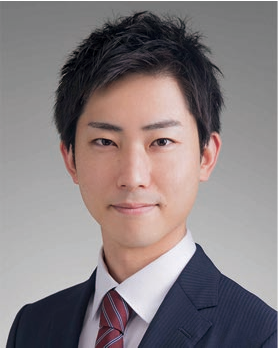}}]{Shuta Kikuchi}
    received the B.~Eng. and M.~Eng. degrees from Waseda University, Tokyo, Japan, in 2017 and 2019, and Dr.~Eng. degrees from Keio University, Kanagawa, Japan in 2024.
    He is currently a Project Assistant Professor with the Graduate School of Science and Technology, Keio University.
    His research interests include Ising machine, statistical mechanics, and quantum annealing.
    He is a member of the Physical Society of Japan (JPS).
\end{IEEEbiography}

\begin{IEEEbiography}[{\includegraphics[width=1in,height=1.25in,clip,keepaspectratio]{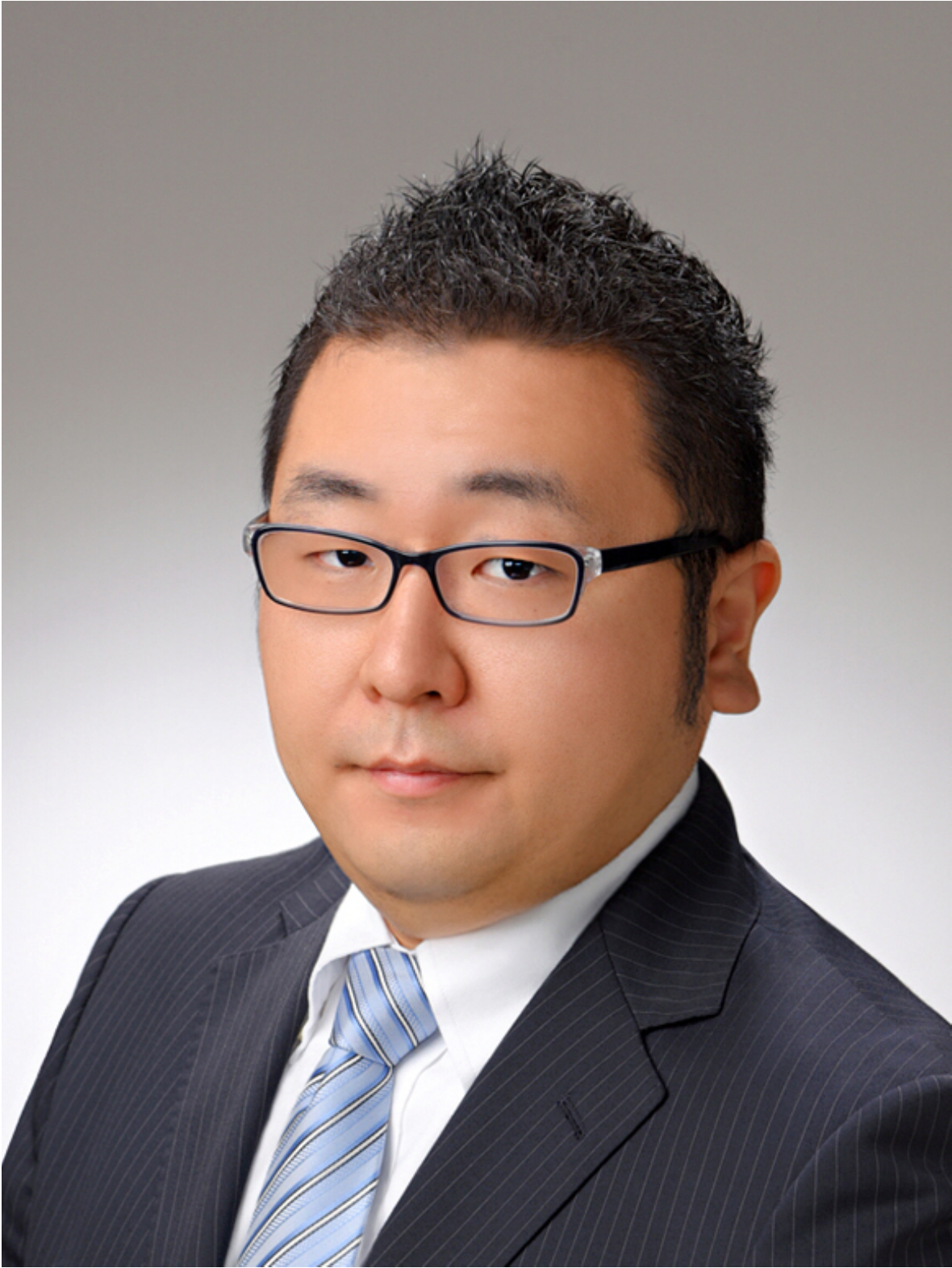}}]{Shu Tanaka} (Member, IEEE)
    received the B.~Sci. degree from the Tokyo Institute of Technology, Tokyo, Japan, in 2003, and the M.~Sci. and Dr.~Sci. degrees from the University of Tokyo, Tokyo, Japan, in 2005 and 2008, respectively.
    He is currently a Professor in the Department of Applied Physics and Physico-Informatics, Keio University, a chair of the Keio University Sustainable Quantum Artificial Intelligence Center (KSQAIC), Keio University, and a Core Director at the Human Biology-Microbiome-Quantum Research Center (Bio2Q), Keio University.
    His research interests include quantum annealing, Ising machines, quantum computing, statistical mechanics, and materials science.
    He is a member of the Physical Society of Japan (JPS), and the Information Processing Society of Japan (IPSJ).
\end{IEEEbiography}

\EOD

\end{document}

%% file: sections/introduction/introduction.tex
\PARstart{C}{ombinatorial} optimization problems (COPs) are optimization problems that aim to find a combination of discrete variables that minimizes an objective function subject to given constraints.
There are many types of COPs and applications in the real world, such as materials design~\cite{Sakaguchi2016,Kitai2020,Endo2022,Inoue2022,couzinie2025machine,xu2025quantum}, portfolio optimization~\cite{Rosenberg2016,Tanahashi2019,Tatsumura2023}, transportation~\cite{Inoue2021,Bao2021,Bao2021-2,Bao2023,kanai2024annealing}, and quantum compilation~\cite{Naito2023}.
Many COPs are classified as NP-hard or NP-complete problems~\cite{Karp2010} and their solution space grows exponentially with the number of variables, making them difficult to solve in polynomial time.
Against this difficulty, Ising machines have attracted attention as a promising metaheuristic approach~\cite{Mohseni2022,Yulianti2022,Jiang2023,Kim2025}.

Ising machines are designed to efficiently search for low-energy states of the Ising Hamiltonian.
Representative approaches include simulated annealing (SA), quantum annealing (QA), simulated quantum annealing (SQA), and coherent Ising machines (CIMs).
SA is a classical stochastic method that mimics thermal fluctuations to escape local minima~\cite{Kirkpatrick1983,Johnson1989,Johnson1991,Isakov2015}.
QA, on the other hand, exploits quantum fluctuations to explore the solution space, and has been implemented on dedicated hardware such as D-Wave’s quantum annealers~\cite{Kadowaki1998,Farhi2000,Farhi2000-2,Johnson2011,tanaka2017quantum,chakrabarti2023quantum}.
QA may provide a quantum speedup for certain classes of problems~\cite{Kadowaki1998,Heim2015,Muthukrishnan2016,Kim2025}.
SQA is a classical algorithm inspired by QA, which approximates quantum tunneling effects using Monte Carlo techniques~\cite{Martonak2002,Kurihara2009,Okuyama2017}.
In addition, CIMs utilize networks of degenerate optical parametric oscillators to naturally evolve toward low-energy configurations of the Ising model~\cite{inagaki2016coherent,Honjo2021}.

To solve COPs using Ising machines, the problems must be mapped to an Ising model whose minimum-energy state corresponds to the optimal solution, or quadratic unconstrained binary optimization (QUBO) formulation~\cite{Lucas2014}.
Such formulations have been developed for many COPs~\cite{Lucas2014,yarkoni2022quantum}.
The Ising model and QUBO are mathematically equivalent; in this paper, we adopt the QUBO formulation.
The QUBO is expressed as
\begin{equation}\label{eq:qubo}
    \mathcal{H}(\bm{x}) = \sum_{i=1}^{N} a_i x_i + \sum_{i=1}^{N-1} \sum_{j=i+1}^{N} b_{ij} x_i x_j,
\end{equation}
where \( \bm{x} = (x_1, x_2, \ldots, x_N)^\top \in \{0, 1\}^N \) is the binary variable vector, \( a_i, b_{ij} \in \mathbb{R} \) are the coefficients, and \( N \) denotes the number of variables.
In practice, real-world COPs often involve multiple constraints simultaneously, and such problems are referred to as constrained combinatorial optimization problems (CCOPs).
For QUBO-based Ising machines, the constraints are typically limited to linear equality and linear inequality constraints on binary variables~\cite{Lucas2014,Glover2022}.
To incorporate such constraints into the QUBO framework, one of the standard approaches is the penalty function method~\cite{Lucas2014}.
Accordingly, the formulation for CCOPs is given as
\begin{equation}\label{eq:ccop}
    \mathcal{H}(\bm x)
    = \mathcal{H}_{\mathrm{obj}}(\bm x)
    + \sum_{m=1}^{M} \mu_m \mathcal{H}_{\mathrm{const}}^{(m)}(\bm x),
\end{equation}
where \( \mathcal{H}_{\mathrm{obj}}(\bm x) \) is the objective function, and
\( \mathcal{H}_{\mathrm{const}}^{(m)}(\bm x) \) represents the \(m\)-th constraint function.
Here, \( \mu_m > 0 \) denotes the penalty coefficient associated with the \(m\)-th constraint, and \( M \) is the total number of constraints.
The constraint function is formulated to take 0 when constraints are satisfied, while a positive value when violated, so that any infeasible solutions become higher energy than the best feasible solution.
Accordingly, we define a solution as feasible if it satisfies all the constraints in~\eqref{eq:ccop}, otherwise it is infeasible.

One of the difficulties when solving the CCOPs with Ising machines is the adjustment of penalty coefficients.
The penalty coefficients in \eqref{eq:ccop} are key parameters that control the trade-off between optimality and feasibility.
The penalty coefficients must be sufficiently large enough to ensure that the minimum-energy state of \( \mathcal{H} \) corresponds to a feasible solution.
However, excessively large penalty coefficients may overemphasize constraint satisfaction, thereby suppressing improvements in the objective function and leading to suboptimal solutions~\cite{Lucas2014,takehara2019multiple,Ayodele2022,Yin2024,Liu2025}.

Another difficulty is the problem size.
Ising machines have hardware limitations such as the number of variables and couplings~\cite{Choi2008,Choi2011}.
Moreover, it has been reported that increasing the number of variables tends to degrade the objective function value~\cite{Mohseni2022,Si2024}.
For this difficulty, the variable reduction is one of the most effective approaches~\cite{Jiang2008,Zou2005,Zhang2005,Shirai2023,Karimi2017,Kikuchi2025,Hattori2025}.
One such method is the sample persistence variable reduction (SPVAR)~\cite{Karimi2017}.
In the SPVAR algorithm, the variables that frequently take the same value in multiple solutions are regarded as the persistent variables, and fixed to that value.
In the previous studies~\cite{Karimi2017,Karimi2017-2}, the SPVAR algorithm has been applied to the 2D and 3D Ising models, weak--strong cluster problems, Chimera graph problems, fault diagnosis problems, and maximum satisfiability problems, and it has been shown to be effective across these benchmark instances in terms of improving standard performance metrics such as the fraction of problems solved, the energy gap to the optimum, the residual error, and the time-to-solution.

Fixing the variables with the SPVAR algorithm involves a finite risk of fixing them to incorrect values, because the SPVAR algorithm is a heuristic variable reduction method.
If variables are incorrectly fixed, the SPVAR algorithm may exclude the optimal solution from the search space.
To mitigate this risk, iterative SPVAR and multi-start SPVAR have been proposed~\cite{Karimi2017,Karimi2017-2}.
There are also several studies about algorithms for Ising machines inspired by the SPVAR algorithm~\cite{Atobe2021,Ayanzadeh2021,Irie2021,Noguchi2023,Kikuchi2023,Lee2024}.

While several extensions of SPVAR have been investigated, these methods have so far been applied to CCOPs with a fixed penalty coefficient.
To the best of our knowledge, no prior work has investigated variable reduction based on sample persistence across solutions obtained with diverse penalty coefficients.
Addressing this gap, in our previous work~\cite{Ide2025}, we proposed the multi-penalty sample persistence variable reduction (MP-SPVAR) algorithm, which explicitly incorporates penalty coefficient diversity into the variable-fixing process.
That study, however, focused primarily on presenting the algorithm itself and reporting preliminary results on a single quadratic assignment problem (QAP) instance.
Moreover, the impact of diverse penalty coefficients on variable persistence was not systematically analyzed.
In this work, we extend the evaluation of MP-SPVAR by applying it to multiple benchmark problems, including both the QAP and the quadratic knapsack problem (QKP), and analyze the low-energy states that emerge under small penalty coefficients.

The rest of this paper is organized as follows.
In Section~\ref{sec:method}, we describe the proposed method in detail, including a review of the SPVAR algorithm, the introduction of the MP-SPVAR algorithm, and the benchmark problems.
In Section~\ref{sec:experimental-setup}, we present the experimental setup.
In Section~\ref{sec:result}, we report the results of the proposed method on the benchmark problems and also evaluate the fixing accuracy of SPVAR and MP-SPVAR.
In Section~\ref{sec:discussion}, we provide a discussion of the performance of the algorithm, focusing on the regime of small penalty coefficients.
Finally, in Section~\ref{sec:conclusion}, we conclude the paper.

%% file: sections/method/method.tex
In this section, we present the details of the proposed approach.
We first review the SPVAR algorithm in Section~\ref{subsec:spvar}, and then introduce our proposed MP-SPVAR algorithm in Section~\ref{subsec:mp-spvar}.
Finally, the benchmark problems used to evaluate the performance of the proposed method are described in Section~\ref{subsec:benchmark}.

\subsection{SPVAR}\label{subsec:spvar}
The SPVAR algorithm was introduced by Karimi and Rosenberg~\cite{Karimi2017} as a variable reduction technique for QUBO formulations.
The central idea of SPVAR is based on the observation that, when a sampler such as a quantum annealer generates a set of low-energy solutions, many variables exhibit persistent values across the sample.
These persistent variables are conjectured to take the same values in the optimal solution.
By fixing such variables to their observed values, the effective problem size can be substantially reduced.

The procedure of the SPVAR algorithm is presented in Algorithm~\ref{alg:spvar}.
Here, \( N_p \) denotes the number of solutions sampled from the sampler,
\( fixing\_threshold \) represents the threshold used to determine whether variables should be fixed,
\( elite\_threshold \) specifies the percentile used to select the top solutions,
and \( sampler \) denotes the solver employed to generate the solutions.
According to the specified \( fixing\_threshold \), variables are fixed to the corresponding binary values (0 or 1).
After fixing the variables, the QUBO is reduced by substituting the fixed values into \eqref{eq:qubo}.
The constant terms are absorbed into the energy offset, while the linear and quadratic coefficients involving the fixed variables are updated accordingly. The resulting reduced QUBO is then solved by the sampler.
It is worth noting that the SPVAR algorithm was shown to be effective with a wide range of solvers, including SA, SQA, and QA~\cite{Karimi2017-2}.

\begin{algorithm}[t]
    \caption{SPVAR}\label{alg:spvar}
    \begin{algorithmic}[1]
        \Require \( N_p \), \( fixing\_threshold \), \( elite\_threshold \), \( sampler \)
        \State Obtain \( N_p \) solutions from \( sampler \).
        \State Retain the top \( elite\_threshold \) percentile solutions.
        \State Calculate the mean value of each variable across the retained solutions.
        \State Fix variables to the corresponding binary values according to the specified \( fixing\_threshold \).
    \end{algorithmic}
\end{algorithm}

\subsection{MP-SPVAR}\label{subsec:mp-spvar}
The MP-SPVAR algorithm was first proposed by Ide, Kikuchi, and Tanaka~\cite{Ide2025}.
The motivation behind the MP-SPVAR algorithm is to extend the SPVAR algorithm to CCOPs.
In this approach, variables that consistently take the same value across multiple solutions obtained with different penalty coefficients are regarded as persistent variables and fixed to that value.
It is expected that solving the reduced problem yields high-quality solutions with high probability, while alleviating the trade-off between feasibility and optimality.

\begin{figure}[t]
    \centering
    \includegraphics[width=0.9\linewidth]{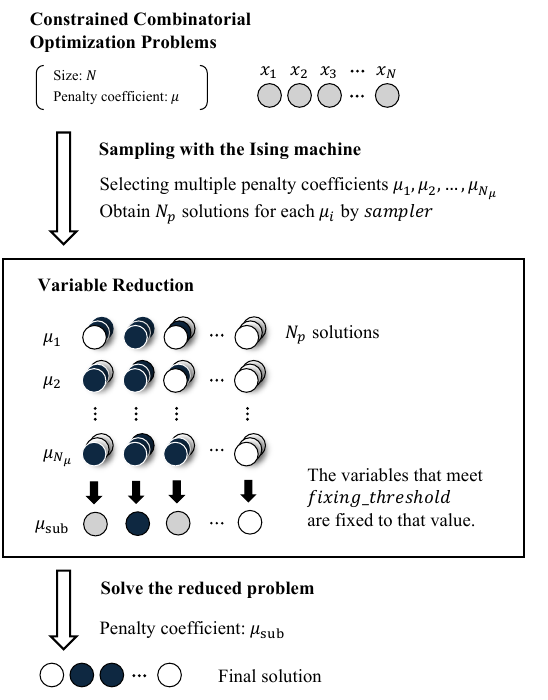}
    \caption{
        Illustration of the MP-SPVAR algorithm.
        The black, white circles represent the variable that takes the value 1 and 0, respectively, and the gray circle represents the undetermined variable.
    }
    \label{fig:mp-spvar}
\end{figure}

The procedure of the MP-SPVAR algorithm is illustrated in Fig.~\ref{fig:mp-spvar}.
First, \( N_\mu \) penalty coefficients \( \left\{ \mu_1, \mu_2, \ldots, \mu_{N_\mu} \right\} \) are chosen to cover a wide range of values to capture diverse solution characteristics.
For each penalty coefficient \( \mu_i \), the CCOP is solved to obtain \( N_p \) solutions.
In total, this process yields \( N_\mu \times N_p \) solutions.
The average solution is then computed by averaging each variable over all collected solutions.
For each variable, if its average value across these solutions satisfies the \( fixing\_threshold \), that variable is fixed to the corresponding binary value (0 or 1).
Finally, the reduced problem is solved using the penalty coefficient that yielded the best feasible objective among the collected solutions.
Note that when \( N_\mu = 1 \), the MP-SPVAR algorithm reduces to the SPVAR algorithm, whereas \( N_\mu = N_p = 1 \) corresponds to the naive method.
Although the $elite\_threshold$ parameter in the SPVAR algorithm is not introduced within the parameter range of this study, it can be incorporated when the number of obtained solutions increases, in order to reduce the number of solutions used for variable fixing.






\subsection{Benchmark Problems}\label{subsec:benchmark}
We apply the proposed method to the representative CCOPs, specifically the quadratic assignment problems (QAP) and quadratic knapsack problem (QKP), as benchmark problems.
In what follows, we present the QUBO formulations of the QAP and the QKP.

The QAP is a CCOP with the equality constraint~\cite{Koopmans1957}, and is known to be NP-hard~\cite{Sahni1976}.
The objective is to assign \( L \) facilities to \( L \) locations, minimizing the total cost of the assignment.
The objective function is given by
\begin{equation}
    \mathcal{H}_{\mathrm{obj}} = \sum_{i=1}^{L} \sum_{j=1}^{L} \sum_{k=1}^{L} \sum_{l=1}^{L} d_{ij} f_{kl} x_{ik} x_{jl},
\end{equation}
where \( d_{ij} \) is the distance between location \( i \) and location \( j \), and \( f_{kl}  \) is the flow from facility \( k \) to facility \( l \).
Throughout, we assume that \( d_{ij} > 0 \) for \( i \neq j \) (with \( d_{ii} = 0 \)) and \( f_{kl} \geq 0 \) for all \( k, l \).
Here, \( x_{ik} \in \{0, 1\} \) is a binary decision variable that takes \( 1 \) if and only if facility \( k \) is assigned to location \( i \), otherwise \( 0 \).
Therefore, when facility \( k \) is assigned to location \( i \) and facility \( l \) to location \( j \), the variables satisfy \( x_{ik} = x_{jl} = 1 \), and the assignment cost \( d_{ij} f_{kl} \) is added to the objective function.
The constraints enforce that each facility is assigned to exactly one location:
\begin{equation}
    \mathcal{H}_{\mathrm{const}} = \sum_{i=1}^{L} \left( \sum_{k=1}^{L} x_{ik} - 1 \right)^2 + \sum_{k=1}^{L} \left( \sum_{i=1}^{L} x_{ik} - 1 \right)^2.
\end{equation}
This constraint is referred to as a two-way one-hot constraint.

The QKP is a CCOP whose objective is to select a subset of items that maximizes the total profit while ensuring that the total weight does not exceed the knapsack capacity.
The problem is known to be NP-hard~\cite{Galli2025}.
The QKP originates from practical applications, such as freight terminal installation~\cite{Rhys1970}, portfolio selection with correlated investments~\cite{Laughhunn1970}, and satellite station placement~\cite{Witzgall1975}.
More recently, QKP has been applied to compiler design~\cite{Johnson1993} and VLSI partitioning~\cite{Ferreira1996}, as well as modern engineering problems such as wind farm layout optimization~\cite{Fischetti2019,Cazzaro2022}.

The QKP with \( N_{\mathrm{item}}  \) items is formulated as
\begin{equation}
    \mathcal{H}_{\mathrm{obj}} = - \sum_{i=1}^{N_{\mathrm{item}}} \sum_{j=i}^{N_{\mathrm{item}}} p_{ij} x_i x_j,
\end{equation}
where \( p_{ii} \geq 0 \) denotes the profit of selecting item \( i \) individually, \( p_{ij} \geq 0 \) denotes the profit obtained by jointly selecting items \( i \) and \( j \) for \( i \neq j \), and \( x_i \in \{0,1\} \) denotes a binary decision variable that equals \( 1 \) if item \( i \) is selected and \( 0 \) otherwise.
The weight constraint is given by
\begin{equation}
    \sum_{i=1}^{N_{\mathrm{item}}} w_i x_i \leq W,
\end{equation}
where \( w_i > 0 \) denotes the weight of item \( i \) and \( W > 0 \) denotes the knapsack capacity.
This inequality can be transformed into an equality by introducing a nonnegative integer slack variable \( S \):
\begin{equation}
    \sum_{i=1}^{N_{\mathrm{item}}} w_i x_i - W + S = 0.
\end{equation}
To embed \( S \) into an Ising machine, binary-integer encodings are employed.
We consider two standard methods: one-hot and domain-wall encoding.

\textbf{One-hot encoding:}
An integer is represented by the position of a single one in a binary vector \(\bm{y} = (y_0, \ldots, y_W) \in \{0, 1\}^{W+1} \).
The encoding constraint and representation of \( S \) are given by
\begin{equation}
    \sum_{k=0}^{W} y_k = 1, \qquad
    S(\bm{y}) = \sum_{k=0}^{W} k y_k.
\end{equation}

\textbf{Domain-wall encoding:}
An integer is represented as a sequence of consecutive ones followed by zeros in \(\bm{y} = (y_0, \ldots, y_{W-1}) \in \{0,1\}^{W} \).
The constraints and the representation of \( S \) are given by
\begin{equation}
    \sum_{k=0}^{W-2} y_{k+1}(1-y_k) = 0, \qquad
    S(\bm{y}) = \sum_{k=0}^{W-1} y_k.
\end{equation}

The constraint Hamiltonian is constructed as
\begin{equation}
    \mathcal{H}_{\mathrm{const}} =
    \lambda \mathcal{H}_{\mathrm{const,weight}}
    + (1-\lambda)\mathcal{H}_{\mathrm{const,encoding}},
\end{equation}
where the relative penalty coefficient \( \lambda \in (0,1) \) balances the weight and encoding penalties and is usually set to \( \lambda = 1/2 \).
For one-hot encoding:
\begin{equation}
    \begin{aligned}
        \mathcal{H}_{\mathrm{const,weight}}^{\mathrm{one-hot}}   & =  \left(\sum_{i=1}^{N_{\mathrm{item}}} w_i x_i - W + \sum_{k=0}^{W} k y_k \right)^2, \\
        \mathcal{H}_{\mathrm{const,encoding}}^{\mathrm{one-hot}} & = \left(\sum_{k=0}^{W} y_k - 1\right)^2,
    \end{aligned}
\end{equation}
and for domain-wall encoding:
\begin{equation}
    \begin{aligned}
        \mathcal{H}_{\mathrm{const,weight}}^{\mathrm{domain-wall}}   & = \left(\sum_{i=1}^{N_{\mathrm{item}}} w_i x_i - W + \sum_{k=0}^{W-1} y_k \right)^2, \\
        \mathcal{H}_{\mathrm{const,encoding}}^{\mathrm{domain-wall}} & = \sum_{k=0}^{W-2} y_{k+1}(1-y_k)
    \end{aligned}
\end{equation}
While the one-hot encoding is widely used~\cite{Chancellor2019,Chen2021,Codognet2022,Zaborniak2023,Ayodele2022}, the domain-wall encoding is known to be more efficient when the relative penalty coefficient \( \lambda \) is properly tuned~\cite{Kikuchi2024}.

%% file: sections/experimental-setup/experimental-setup.tex
All experiments were conducted using the Fixstars Amplify Annealing Engine~\cite{FixstarsAmplify2025}, which is a graphics processing unit (GPU)-based Ising machine.
The computational time can be specified by the user and was set to \( \SI{5000}{ms} \) unless otherwise noted.

QAP instances were selected from QAPLIB~\cite{Burkard1997}, which provides a standard benchmark set for QAP.
The library contains more than one hundred instances, including real-world cases (such as hospital layouts, backboard wiring, and keyboard design), randomly generated problems, and structured synthetic instances based on grid distances or algebraic constructions.
The problem sizes range from small instances with \( L \approx 10 \)  (i.e., on the order of a few hundred variables  in QUBO form), where optimal solutions are known, to large instances with \( L \geq 100 \) (i.e., tens of thousands of binary variables), for which only heuristic solutions and lower bounds are available.
The distance matrix \( d_{ij} \) and flow matrix \( f_{kl} \) were normalized by dividing each entry by the maximum element of the respective matrix.
From this benchmark set, we used tai40b, tai60b, tai80b, and tai100b for the following experiments.

QKP instances were selected from the QKP Instance library~\cite{Billionnet2004,Billionnet2004-2}, which provides a standard set of benchmark instances for the QKP.
Each instance is characterized by two parameters: the number of items \( n \in \{100, 200, 300\} \) and the density of the profit matrix \( d \in \{25\%, 50\%, 75\%, 100\% \} \).
For each \((n,d)\) pair, ten instances are provided, except for the cases \((300,75\%)\) and \((300,100\%)\), where some instances are missing.
The optimal value for all instances is known.
To facilitate comparison across different instances, the objective function \(\mathcal{H}_{\mathrm{obj}}\) was normalized by dividing all profit entries by the maximum value of the profit matrix in each instance.
In this study, we used the instance \textit{r\_100\_25\_6}, which is the sixth instance for \((n,d) = (100,25\%)\).

For the hyperparameters of the MP-SPVAR algorithm, the following settings were used.
For the \( fixing\_threshold \), only variables that took identical values across all sampled solutions were fixed.
The number of solutions \( N_p \) was varied from 1 to 10, and the combinations of penalty coefficients \( \mu \) were chosen from the set
\(\{0.01, 0.1, 1, 10, 100, 1000\}\).
Since the objective functions of both the QAP and QKP were normalized, we consider the selected range of penalty coefficients to adequately cover values from sufficiently small to sufficiently large.
Each computation was performed 10 times for every combination of penalty coefficients and \( N_p \) .
The details of the dependencies of the optimality and feasibility on the penalty coefficients are provided in Appendix~\ref{sec:appendix-naive}.

%% file: sections/result/result.tex
In this section, we present the experimental results.
In Section~\ref{subsec:result-spvar}, we analyze the dependency of the performance of the SPVAR algorithm on the penalty coefficient.
In Section~\ref{subsec:result-mp-spvar}, we report the performance of the proposed MP-SPVAR algorithm on benchmark problems.
In Section~\ref{subsec:result-fixing-accuracy}, we investigate the fixing accuracy of both SPVAR and MP-SPVAR algorithms.

\subsection{Penalty coefficient dependency of SPVAR algorithm}\label{subsec:result-spvar}
First, we evaluated the performance of the SPVAR algorithm on the CCOPs, especially penalty coefficient dependency.

In this analysis, we employed two evaluation metrics: the approximation ratio (approx.\ ratio) and the feasible-solution ratio (FS\ ratio).
The approx.\ ratio is defined as the ratio of the feasible objective value to the best-known optimal value:
\begin{equation}\label{eq:approx-ratio}
    \mathrm{Approx.\ Ratio} =
    \frac{\text{Feasible Objective Value}}{\text{Best-Known Optimal Value}}.
\end{equation}
For maximization problems, the value of the approx.\ ratio lies between 0 and 1,
whereas for minimization problems, it takes values greater than 1.
In both cases, values closer to 1 indicate better solution quality.
The best-known optimal values were taken from the corresponding benchmark problem sets described in Section~\ref{sec:experimental-setup}.
The approx.\ ratio was computed only for runs in which a feasible solution was obtained, using the feasible objective value from that run.
The reported values represent the average of such runs, excluding runs without feasible solutions.
The FS\ ratio measures the proportion of feasible solutions among the runs for a given combination of \(\mu\) and \(N_p\):
\begin{equation}\label{eq:fs-ratio}
    \mathrm{FS\ Ratio} =
    \frac{\text{Number of Feasible Solutions}}{\text{Total Number of Runs}}.
\end{equation}
Specifically, the numerator in \eqref{eq:fs-ratio} counts the feasible solutions obtained, and the denominator in \eqref{eq:fs-ratio} corresponds to the total number of runs.
For the QAP, feasibility requires that the assignment constraints are strictly satisfied; namely, each facility must be assigned to exactly one location, and each location must be occupied by exactly one facility.
For the QKP, feasibility is defined by the simultaneous satisfaction of the knapsack weight constraint and the encoding constraint imposed by the slack variable.

\begin{figure}[t]
    \centering
    \begin{minipage}[ht]{0.48\hsize}\centering
        \includegraphics[width=\linewidth]{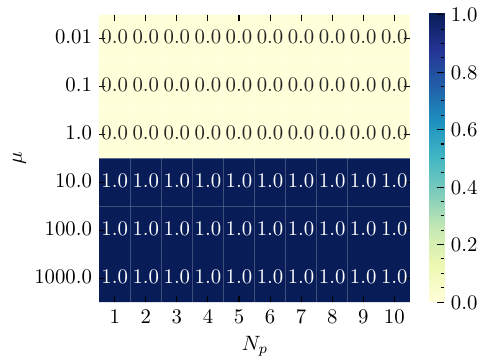}
        \subcaption{}\label{subfig:spvar-qap-feasible-tai60b}
    \end{minipage}
    \begin{minipage}[ht]{0.48\hsize}\centering
        \includegraphics[width=\linewidth]{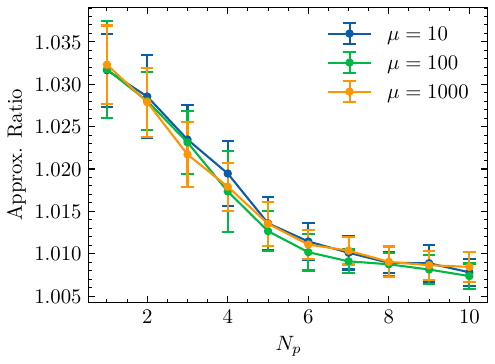}
        \subcaption{}\label{subfig:spvar-qap-approx-tai60b}
    \end{minipage}
    \caption{
        (a) FS\ ratio and (b) approx.\ ratio with respect to the number of solutions \(N_p\) for the 60 facility QAP using the SPVAR algorithm.
        Here, \(N_p\) denotes the number of solutions, and variables that take identical values across all solutions were fixed accordingly.
        The penalty coefficient \(\mu\) was chosen from the set \(\{0.01, 0.1, 1, 10, 100, 1000\}\).
        Each process was performed 10 times.
        Note that in panel (b), the curves for \(\mu = 0.01, 0.1,\) and \(1\) are omitted because these penalty coefficients did not yield any feasible solutions.
    }\label{fig:spvar-qap-tai60b}
\end{figure}

Fig.~\ref{fig:spvar-qap-tai60b} shows the (a) FS\ ratio and (b) approx.\ ratio with respect to the number of solutions \(N_p\) for the 60-facility QAP instance tai60b using the SPVAR algorithm.
The penalty coefficient \(\mu\) was selected from the set \(\{0.01, 0.1, 1, 10, 100, 1000\}\), and each experiment was performed 10 times.
As \(N_p\) increased, the approx.\ ratio improved, whereas the FS\ ratio remained unchanged, as it was determined solely by the value of \(\mu\).
When \(\mu\) was so small that the naive (\(N_p = 1\)) method failed to yield a feasible solution, the SPVAR algorithm likewise produced no feasible solution, as discussed in Section~\ref{subsec:discussion-spvar-mp-spvar-variable-fixing} where the reasons are considered.
Conversely, when \(\mu\) is sufficiently large such that the naive method  obtains a feasible solution, the SPVAR algorithm also succeeds, for the same reasons discussed in Section~\ref{subsec:discussion-spvar-mp-spvar-variable-fixing}.
This trend is consistently observed across other QAP instances, with detailed results provided in Appendix~\ref{sec:appendix-spvar}.
These findings indicate that careful tuning of the penalty coefficient is crucial when applying the SPVAR algorithm to QAPs.

\begin{figure}[t]
    \centering
    \begin{minipage}[ht]{0.48\hsize}\centering
        \includegraphics[width=\linewidth]{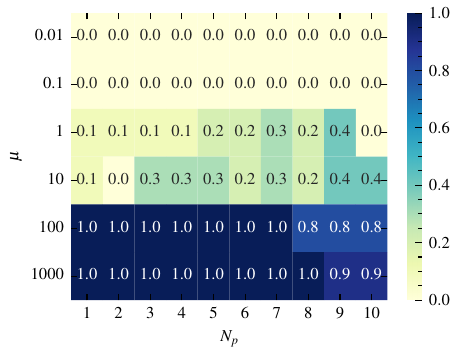}
        \subcaption{}\label{subfig:spvar-qkp-feasible-jeu_100_25_6-onehot}
    \end{minipage}
    \hfill
    \begin{minipage}[ht]{0.48\hsize}\centering
        \includegraphics[width=\linewidth]{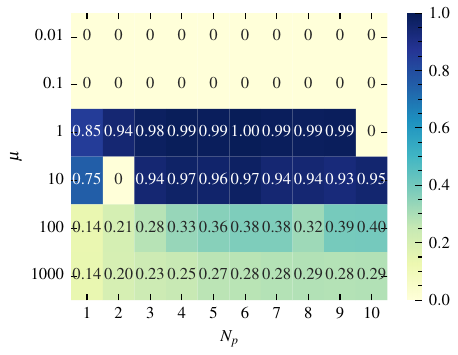}
        \subcaption{}\label{subfig:spvar-qkp-approx-jeu_100_25_6-onehot}
    \end{minipage}
    \begin{minipage}[ht]{0.48\hsize}\centering
        \includegraphics[width=\linewidth]{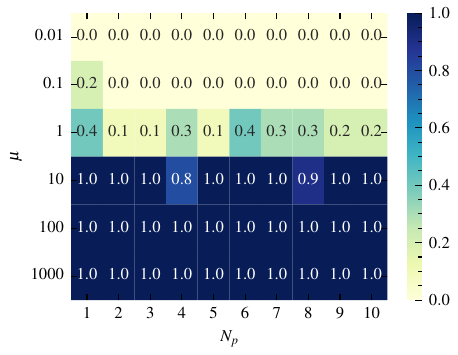}
        \subcaption{}\label{subfig:spvar-qkp-feasible-jeu_100_25_6-domain_wall}
    \end{minipage}
    \hfill
    \begin{minipage}[ht]{0.48\hsize}\centering
        \includegraphics[width=\linewidth]{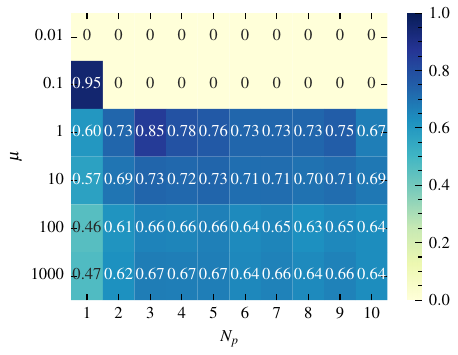}
        \subcaption{}\label{subfig:spvar-qkp-approx-jeu_100_25_6-domain_wall}
    \end{minipage}
    \caption{
        (a), (c) FS\ ratio and (b), (d) approx.\ ratio with respect to the number of solutions \(N_p\) for the 100 items, 1236 capacity QKP instance using the SPVAR algorithm.
        (a), (b) show the results with the one-hot encoding, and (c), (d) show the results with the domain-wall encoding, respectively.
        In (b), (d), the value 0 means that no feasible solution was obtained.
        Here, \(N_p\) denotes the number of solutions, and variables that take identical values across all solutions are fixed accordingly.
        The penalty coefficient \( \lambda \) was set to 0.5 for one-hot encoding, and 0.6 for domain-wall encoding.
        Each process was performed 10 times.
    }\label{fig:spvar-qkp-jeu_100_25_6}
\end{figure}

Fig.~\ref{fig:spvar-qkp-jeu_100_25_6} shows the (a), (c) FS\ ratio and (b), (d) approx.\ ratio with respect to the number of solutions \(N_p\) for the 100 items, 1236 capacity QKP instance r\_100\_25\_6 using the SPVAR algorithm.
Fig.~\ref{fig:spvar-qkp-jeu_100_25_6} (a) and (b) present the results obtained with the one-hot encoding, while Fig.~\ref{fig:spvar-qkp-jeu_100_25_6} (c) and (d) show those with the domain-wall encoding.
The relative penalty coefficient \( \lambda \) was set to \( 0.5 \) for the one-hot encoding and \( 0.6 \) for the domain-wall encoding.
The results for other values of \(\lambda\) are provided in Appendix~\ref{sec:appendix-mp-spvar}.
In the QKP, which is a maximization problem, the approx.\ ratio takes values in the range from 0 to 1.
Similar to the QAP, values closer to \(1\) indicate better performance.
Overall, as in the case of the QAP, when the penalty coefficient was sufficiently large such that a feasible solution was obtained with probability one in the naive method, the SPVAR algorithm also yielded feasible solutions with high probability.
Moreover, when the probability of obtaining a feasible solution was zero in the naive method, it was likewise zero in the SPVAR algorithm.
Unlike the QAP, however, there existed an intermediate range of penalty coefficients that yielded good objective function values with low probability, where the behavior with respect to \( N_p \) differed depending on the encoding.
The trade-off between the approx.\ ratio and the FS\ ratio were more pronounced in the QKP than in the QAP.

For the one-hot encoding, the approx.\ ratio improved as \(N_p\) increased when the penalty coefficient was sufficiently large to consistently yield feasible solutions, specifically for \(\mu = 100, 1000\), as observed in the QAP as well.
When the penalty coefficient produced feasible solutions with probability one in the naive method (\(N_p = 1\)), such as \(\mu = 100\) or \(\mu = 1000\), the SPVAR algorithm also achieved feasible solutions with high probability.
Conversely, when the penalty coefficient produced feasible solutions only with low probability, such as \(\mu = 1\) or \(\mu = 10\), the SPVAR algorithm likewise produced feasible solutions with low probability.
When \(\mu = 0.01\) or \(\mu = 0.1\), which yielded no feasible solutions in the naive method, the SPVAR algorithm also failed to produce feasible solutions.
For these intermediate values of \(\mu\), the FS\ ratio tends to increase as \(N_p\) increases, whereas for larger penalty coefficients the FS\ ratio decreases when \(N_p\) becomes large.

For the domain-wall encoding, when the penalty coefficient was sufficiently large to yield feasible solutions with probability one in the naive method (e.g., \( \mu = 10, 100, 1000 \)), the SPVAR algorithm also produced feasible solutions with high probability.
With such large penalty coefficients, the improvement in the approx.\ ratio with increasing \( N_p \) was limited; however, a clear gain was observed when moving from the naive method to \( N_p \geq 2 \).
When comparing the two encodings, the degradation of the approx.\ ratio with increasing \(\mu\) was more severe for the one-hot encoding than for the domain-wall encoding.
Therefore, as in the case of the QAP, it is necessary to adjust the penalty coefficient appropriately in the QKP in order to obtain feasible and high-quality solutions.

\subsection{MP-SPVAR}\label{subsec:result-mp-spvar}
To evaluate the effectiveness of the MP-SPVAR algorithm, we apply it to the same QAP and QKP instances analyzed in the previous subsection and compare the results with those of the SPVAR algorithm.
As in Section~\ref{subsec:result-spvar}, the evaluation metrics used were the approx.\ ratio and FS\ ratio.

\begin{figure*}[!t]
    \centering
    \begin{minipage}[ht]{0.48\hsize}\centering
        \includegraphics[width=\linewidth]{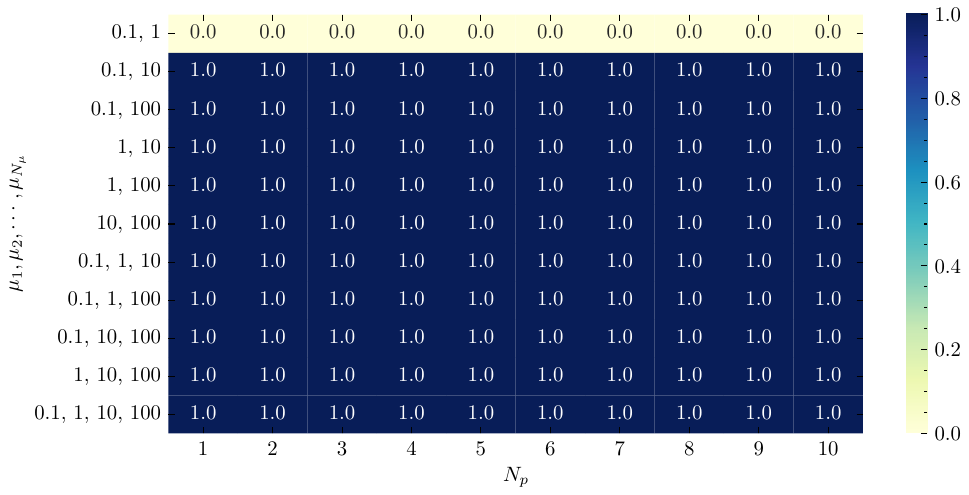}
        \subcaption{}
    \end{minipage}
    \begin{minipage}[ht]{0.48\hsize}\centering
        \includegraphics[width=\linewidth]{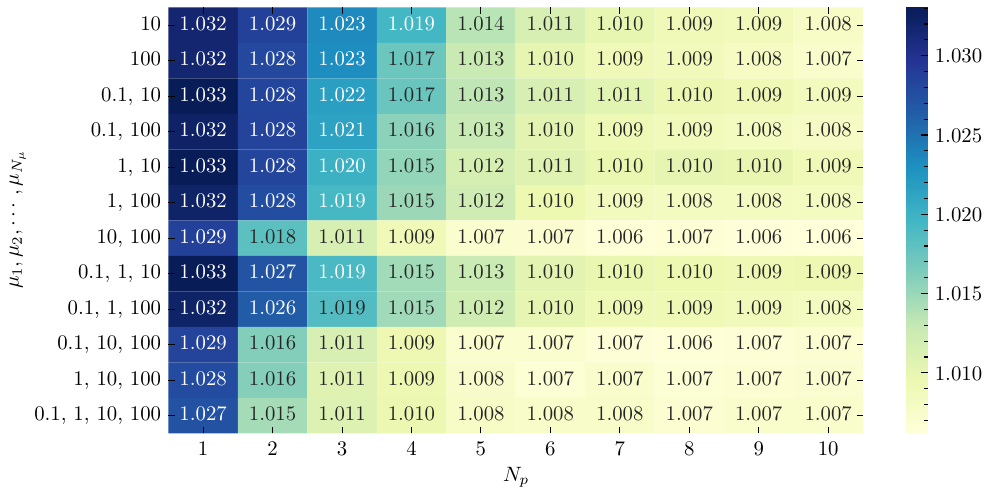}
        \subcaption{}
    \end{minipage}
    \caption{
        (a) FS\ ratio and (b) approx.\ ratio with respect to the number of solutions \(N_p\) for the 60 facility QAP using the MP-SPVAR algorithm.
        \( N_p \) is the number of solutions for each penalty coefficient.
        The penalty coefficients \( \mu_1, \cdots, \mu_{N_\mu} \) were chosen from all the combinations of the set \(\{0.01, 0.1, 1, 10, 100, 1000\}\).
        Each process was performed 10 times.
    }\label{fig:mp-spvar-qap-tai60b}
\end{figure*}

Fig.~\ref{fig:mp-spvar-qap-tai60b} shows the (a) FS\ ratio and (b) approx.\ ratio with respect to the number of solutions \(N_p\) for the 60-facility QAP instance tai60b using the MP-SPVAR algorithm.
All combinations of the set \(\{0.1, 1, 10, 100\}\) were used as the penalty coefficients.
From the preliminary results with the SPVAR method, we observed that smaller coefficients such as 0.1 and 1 consistently produced infeasible solutions, while larger coefficients such as 10 and 100 yielded feasible ones.
By combining these two distinct regimes (infeasible vs. feasible), the chosen set provides a clear contrast that facilitates the evaluation of MP-SPVAR performance across different feasibility conditions.
The MP-SPVAR algorithm achieved high feasibility whenever the set of penalty coefficients included at least one value that yielded feasible solutions in the naive method.
It also maintained solution quality comparable to that of SPVAR, while benefiting from the diversity introduced by multiple penalty coefficients.
The FS\ ratio reached \(1.0\) whenever the penalty coefficient set contained at least one value that produced feasible solutions.
The approx.\ ratio improved with increasing \(N_p\), similar to the SPVAR algorithm.
Moreover, the approx.\ ratio was comparable to that of SPVAR, or slightly smaller.
This held when the value of \(N_p\) times the number of penalty coefficients producing feasible solutions was identical.
For example, when comparing the case of \(N_\mu = 1, \mu = 10\) with that of \(N_\mu = 2, \mu_1 = 10, \mu_2 = 100\), the approx.\ ratio remained nearly the same, or was slightly smaller in the latter case, provided that the product of the number of penalty coefficients yielding feasible solutions and \(N_p\) was identical.
In contrast, even penalty coefficients that predominantly generated infeasible solutions contribute to a slight improvement in the approx.\ ratio.
For instance, when comparing \(N_\mu = 1, \mu = 10\) with \(N_\mu = 3, \mu_1 = 0.1, \mu_2 = 1, \mu_3 = 10\), the inclusion of infeasible solutions also contributed to improving the approx.\ ratio as \(N_p\) increased.
\begin{figure*}[!t]
    \centering
    \begin{minipage}[ht]{0.48\hsize}\centering
        \includegraphics[width=\linewidth]{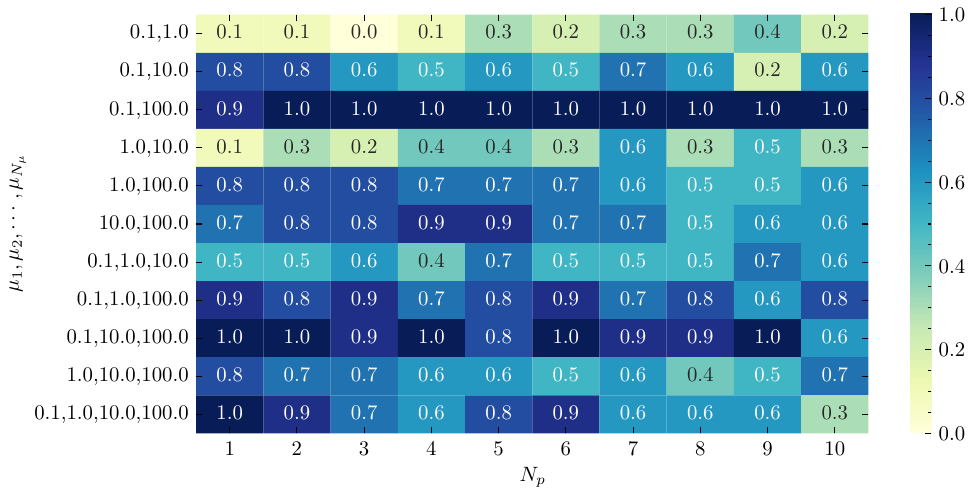}
        \subcaption{}\label{subfig:mp-spvar-qkp-jeu_100_25_6-onehot-fs-ratio}
    \end{minipage}
    \begin{minipage}[ht]{0.48\hsize}\centering
        \includegraphics[width=\linewidth]{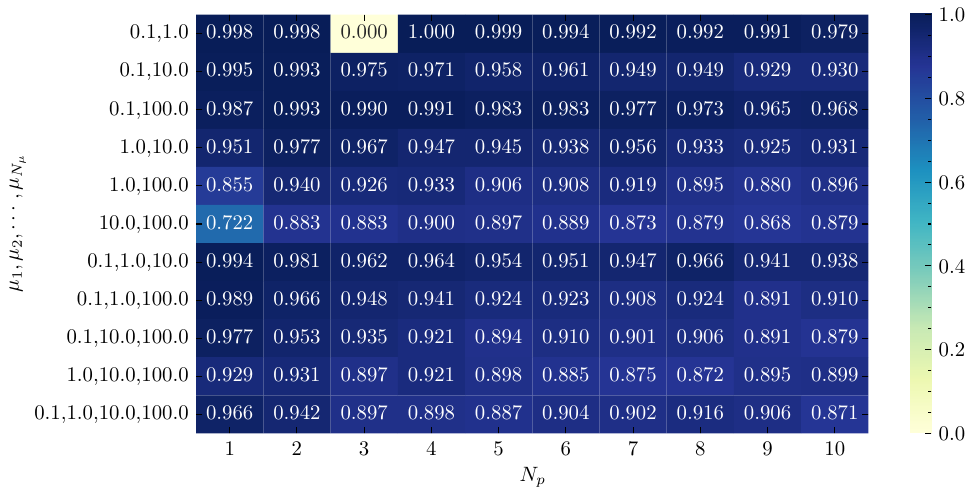}
        \subcaption{}\label{subfig:mp-spvar-qkp-jeu_100_25_6-onehot-approx-ratio}
    \end{minipage}
    \begin{minipage}[ht]{0.48\hsize}\centering
        \includegraphics[width=\linewidth]{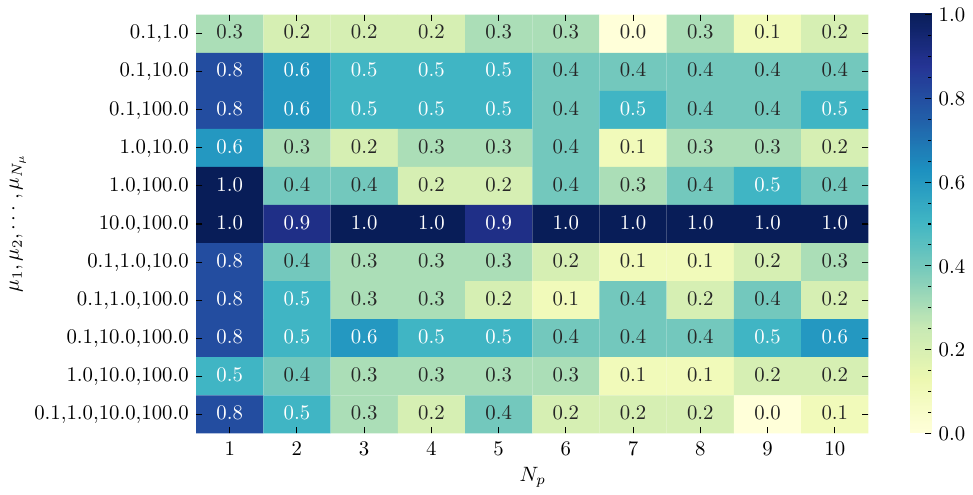}
        \subcaption{}\label{subfig:mp-spvar-qkp-jeu_100_25_6-domain_wall-fs-ratio}
    \end{minipage}
    \begin{minipage}[ht]{0.48\hsize}\centering
        \includegraphics[width=\linewidth]{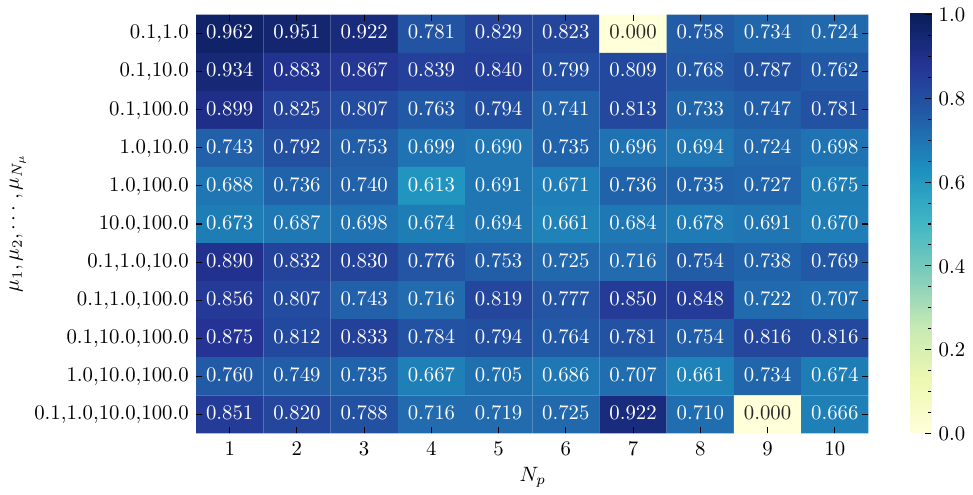}
        \subcaption{}\label{subfig:mp-spvar-qkp-jeu_100_25_6-domain_wall-approx-ratio}
    \end{minipage}
    \caption{
        (a), (c) FS\ ratio and (b), (d) approx.\ ratio with respect to the number of solutions \(N_p\) for the 100 items, 1236 capacity QKP instance using the MP-SPVAR algorithm.
        (a), (b) show the results with the one-hot encoding, and (c), (d) show the results with the domain-wall encoding, respectively.
        In (a), (c), the value 0 means that the no feasible solution was obtained.
        \( N_p \) is the number of solutions for each penalty coefficient.
        The penalty coefficients \( \mu_1, \cdots, \mu_{N_\mu} \) were chosen from all the combinations of the set \(\{0.1, 1, 10, 100\}\).
        Each process was performed 10 times.
    }\label{fig:mp-spvar-qkp-jeu_100_25_6}
\end{figure*}

For the QKP, we conducted experiments analogous to those for the QAP, and the results are shown in Fig.~\ref{fig:mp-spvar-qkp-jeu_100_25_6}, which presents the approx.\ ratio and FS\ ratio with respect to the number of solutions \(N_p\) for the 100 items, 1236 capacity instance using the MP-SPVAR algorithm.
Fig.~\ref{fig:mp-spvar-qkp-jeu_100_25_6} (a) and (b) present the results with one-hot encoding, while Fig.~\ref{fig:mp-spvar-qkp-jeu_100_25_6} (c) and (d) present those with domain-wall encoding.
All combinations of the set \(\{0.1, 1, 10, 100\}\) were used as the penalty coefficients, and the relative penalty coefficient \(\lambda\) was set to \(0.5\) for one-hot encoding and \(0.6\) for domain-wall encoding.
This set was chosen because, as in the QAP, the SPVAR results indicated that \(\mu=0.1\) yielded an FS\ ratio close to zero, whereas \(\mu=100\) yielded an FS\ ratio close to one.
Therefore, the selected set was considered most suitable for evaluating the characteristics of the MP-SPVAR algorithm.
Other results with different values of \(\lambda\) are provided in Appendix~\ref{sec:appendix-mp-spvar}.
For both encodings, the MP-SPVAR algorithm yielded approx.\ ratio and FS\ ratio values that simultaneously reflect the high FS\ ratio typically obtained with larger penalty coefficients and the high approx.\ ratio associated with smaller penalty coefficients.
In contrast to the QAP, no clear trend with respect to \(N_p\) was observed; moreover, in the case of the domain-wall encoding, the performance even deteriorated.

Focusing on the one-hot encoding, feasible solutions were obtained for almost all combinations of penalty coefficients, except for the case of \( N_{\mu} = 2, \mu = 0.1, 1 \) with \( N_p = 3 \), where no feasible solution was found.
No clear trend with respect to \( N_p \) was observed for the FS\ ratio.
In contrast, when the penalty coefficient set included \( \mu = 100 \), which reliably yielded feasible solutions in both the naive method and the SPVAR algorithm, a consistently high FS\ ratio was obtained.
Regarding the approx.\ ratio, it remained high—exceeding 0.8 for nearly all penalty coefficients.
For instance, with the combination \( \mu = 0.1, 100 \), feasible solutions were obtained with a probability greater than 0.9 regardless of \( N_p \), and the approx.\ ratio also exceeded 0.9.

For the domain-wall encoding, the FS\ ratio showed no clear trend with respect to increasing \( N_p \), similar to the one-hot encoding, but tended to deteriorate slightly, with \( N_p = 1 \) often yielding the best values for a given combination of penalty coefficients.
Regarding the choice of penalty coefficients, as in the one-hot encoding, the FS ratio remained high for most combinations, except for the case \(N_\mu=2\) with \(\mu_1=0.1\) and \(\mu_2=1\), which did not yield feasible solutions under either the naive method or the SPVAR algorithm.
Notably, no feasible solutions were obtained for the combinations \(\mu = 0.1, 1\) with \( N_p = 7 \) and \(\mu = 0.1, 1, 10, 1000\) with \( N_p = 9 \).
For the approx.\ ratio, no clear dependence on \( N_p \) can be observed.
Across almost all combinations of penalty coefficients, the approx.\ ratio remained at or above 0.613.

\subsection{Fixing Accuracy of the SPVAR and MP-SPVAR Algorithms}\label{subsec:result-fixing-accuracy}
In this subsection, we evaluated which variables were fixed and how accurately they were fixed by the SPVAR and MP-SPVAR algorithms.
Fixing variables to incorrect values makes it impossible to obtain the optimal solution from the reduced problem; therefore, it is essential to avoid any incorrect fixings.
On the other hand, fixing too few variables is insufficient for reaching the optimal solution.
Thus, we assess both the number of fixed variables and the accuracy of their assignments.
A variable is regarded as incorrectly fixed if its fixed value differs from that in the known optimal solution.
For the QAP, we used the optimal solutions provided in the benchmark set.
For the QKP, since the benchmark set does not include the optimal solutions themselves but only their objective values, we regarded as the optimal solution any solution obtained during the experiments whose objective value matched the optimal one.
The optimal solution we obtained is provided in Appendix~\ref{sec:appendix-qkp-optimal-solution}

Fig.~\ref{fig:mp-spvar-qap-tai60b-fixed} shows the fixed ratio and incorrect fix count with respect to the number of solutions \(N_p\) for the 60 facility QAP instance tai60b using the MP-SPVAR algorithm.
The fixed ratio is the ratio of the variables that are fixed to the total number of variables:
\begin{equation}
    \textrm{Fixed Ratio} = \dfrac{\textrm{Number of Fixed Variables}}{\textrm{Number of Variables}},
\end{equation}
and incorrect fix count is the number of variables that are fixed to the incorrect value.
As \( N_p \) increased, both fixed ratio and incorrect fix count decreased.
Within the tested range of \(N_p\), a nonzero incorrect-fix count was observed even under the strictest \( fixing\_threshold \) setting, where only variables that take identical values across all solutions are fixed.
This result indicates that a larger \(N_p\) is required to suppress incorrect variable fixing and to reliably obtain the optimal solution.
It should be noted, however, that this interpretation assumes the sampling characteristics of the annealer are fixed, and that improvements in the annealer itself could also reduce erroneous fixings with fewer samples.

\begin{figure}[t]
    \centering
    \begin{minipage}[ht]{0.48\hsize}\centering
        \includegraphics[width=\linewidth]{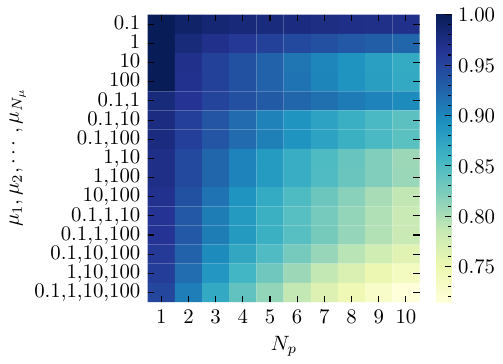}
        \subcaption{}
    \end{minipage}
    \hfill
    \begin{minipage}[ht]{0.48\hsize}\centering
        \includegraphics[width=\linewidth]{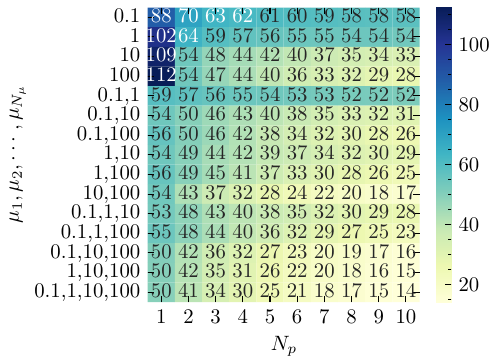}
        \subcaption{}
    \end{minipage}
    \caption{
        The averages of the (a) fixed ratio and (b) incorrect fix count with respect to the number of solutions \(N_p\) for the 60 facility QAP, obtained over 10 runs using the SPVAR and MP-SPVAR algorithm
    }\label{fig:mp-spvar-qap-tai60b-fixed}
\end{figure}

\begin{figure}[t]
    \centering
    \begin{minipage}[ht]{0.48\hsize}\centering
        \includegraphics[width=\linewidth]{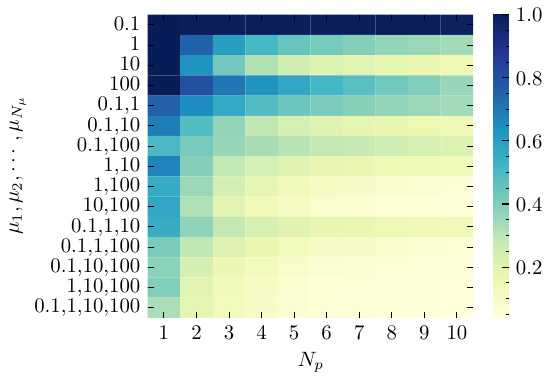}
        \subcaption{}
    \end{minipage}
    \hfill
    \begin{minipage}[ht]{0.48\hsize}\centering
        \includegraphics[width=\linewidth]{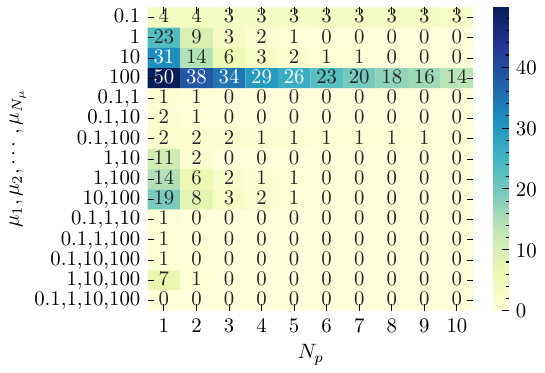}
        \subcaption{}
    \end{minipage}
    \begin{minipage}[ht]{0.48\hsize}\centering
        \includegraphics[width=\linewidth]{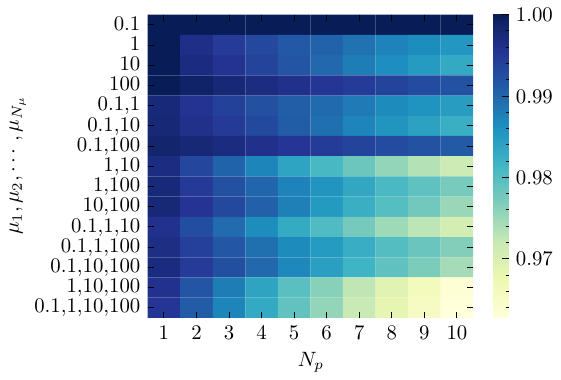}
        \subcaption{}
    \end{minipage}
    \hfill
    \begin{minipage}[ht]{0.48\hsize}\centering
        \includegraphics[width=\linewidth]{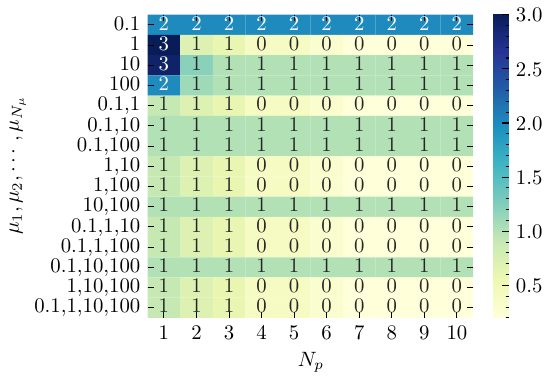}
        \subcaption{}
    \end{minipage}
    \caption{
        (a), (c) Fixed ratio and (b), (d) incorrect fix count with respect to the number of solutions \(N_p\) for the 100 items, 1236 capacity QKP instance with the one-hot encoding using the MP-SPVAR algorithm.
        (a), (b) show the results of decision variables, and (c), (d) show the results of slack variables, respectively.
        \( N_p \) is the number of solutions for each penalty coefficient.
        The penalty coefficients \( \mu_1, \cdots, \mu_{N_\mu} \) were chosen from all the combinations of the set \(\{0.1, 1, 10, 100\}\).
        Each process was performed 10 times.
        The penalty coefficient \( \lambda \) was set to 0.5.
    }\label{fig:mp-spvar-qkp-jeu_100_25_6-fixed-onehot}
\end{figure}

\begin{figure}[!t]
    \centering
    \begin{minipage}[ht]{0.48\hsize}\centering
        \includegraphics[width=\linewidth]{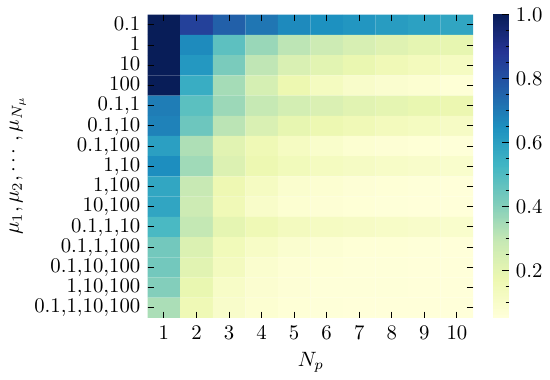}
        \subcaption{}
    \end{minipage}
    \hfill
    \begin{minipage}[ht]{0.48\hsize}\centering
        \includegraphics[width=\linewidth]{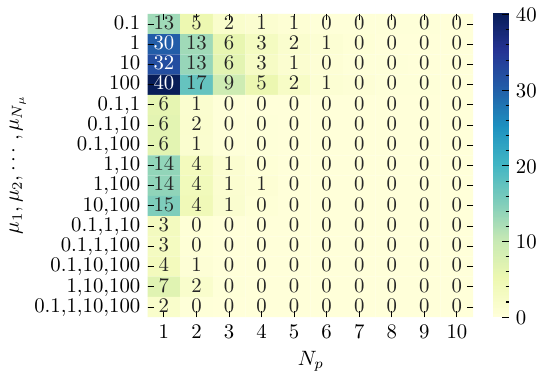}
        \subcaption{}
    \end{minipage}
    \begin{minipage}[ht]{0.48\hsize}\centering
        \includegraphics[width=\linewidth]{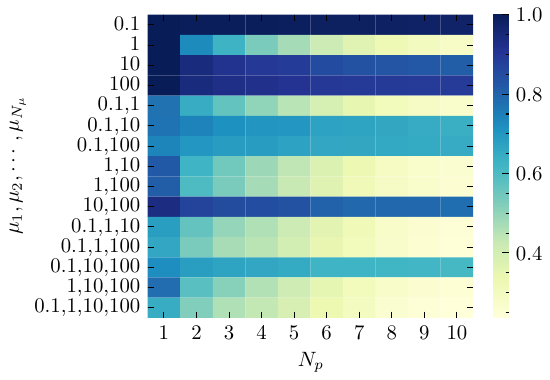}
        \subcaption{}
    \end{minipage}
    \hfill
    \begin{minipage}[ht]{0.48\hsize}\centering
        \includegraphics[width=\linewidth]{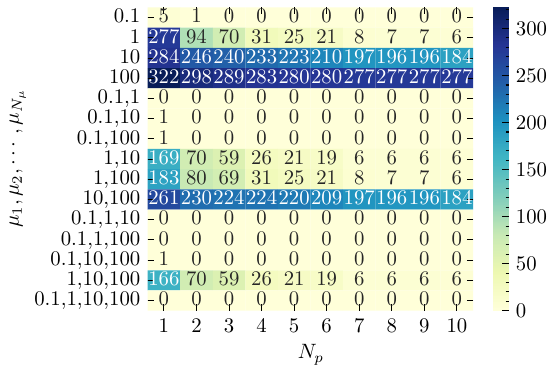}
        \subcaption{}
    \end{minipage}
    \caption{
        (a), (c) Fixed ratio and (b),(d) incorrect fix count with respect to the number of solutions \(N_p\) for the 100 items, 1236 capacity QKP instance with the domain-wall encoding using the MP-SPVAR algorithm.
        (a), (b) show the results of fixed variables, and (c), (d) show the results of slack variables, respectively.
        \( N_p \) is the number of solutions for each penalty coefficient.
        The penalty coefficients \( \mu_1, \cdots, \mu_{N_\mu} \) were chosen from all the combinations of the set \(\{0.1, 1, 10, 100\}\).
        Each process was performed 10 times.
        The penalty coefficient \( \lambda \) was set to 0.6.
    }\label{fig:mp-spvar-qkp-jeu_100_25_6-fixed-domain_wall}
\end{figure}

In the case of the QKP, the fixed ratio and incorrect fix count are shown in Fig.~\ref{fig:mp-spvar-qkp-jeu_100_25_6-fixed-onehot} and \ref{fig:mp-spvar-qkp-jeu_100_25_6-fixed-domain_wall} for the one-hot encoding and the domain-wall encoding, respectively.
In both figures, (a) and (b) show the results of the decision variables \( x_i \), and (c) and (d) show the results of the slack variables \( y_k \).
First, we consider the case of one-hot encoding under the SPVAR algorithm with \( N_\mu = 1 \).
For the slack variables, the ratio of fixed variables remained above 0.95, indicating that almost all of them were fixed.
Even if the one-hot constraint is satisfied with \(N_{\mu} = 4\) and \(N_{p} = 10\) solutions, each yielding a distinct integer, the value remains
\((1237 - 40)/1237 = 0.967\) under the strictest \(fixing\_threshold\).
Thus, when the condition is enforced solely by \(\mu\) values that satisfy the one-hot constraint, the result readily exceeds 0.95.
In contrast, for the case with \(\{0.1, 1, 10, 100\}\) and \(N_{p} = 10\), the value fell below 0.965, suggesting that under small penalty coefficients the one-hot constraint was not satisfied, and multiple slack variables were simultaneously assigned the value 1.
When the penalty coefficient was small (e.g., \(\mu = 0.1\)), almost all decision variables are fixed, and among the decision variables \(x_i\), three or four variables are incorrectly fixed.
Similarly, for the slack variables, nearly all of them were fixed, and two variables were incorrectly fixed.
As the penalty coefficient increased, the incorrect fix count tended to increase.
With increasing \( N_p \), both the number of fixed decision variables and the number of incorrectly fixed variables decreased.
In the case of the MP-SPVAR method, compared with the SPVAR method, both decision and slack variables exhibited lower fixed ratios, and the proportion of incorrectly fixed variables is also reduced.

For the domain-wall encoding, a similar overall trend was observed as in the one-hot encoding.
However, the fixed ratio of slack variables was significantly lower than that of the one-hot encoding.
The proportion of incorrectly fixed slack variables under the SPVAR method was larger than in the one-hot case, but the use of multiple penalty coefficients in the MP-SPVAR method substantially reduced the number of incorrectly fixed variables.
In particular, when the penalty coefficient set included \(\mu = 0.1\), the proportion of incorrectly fixed variables became nearly zero.

%% file: sections/discussion/discussion.tex
In this section, we analyze the low-energy states at small and large penalty coefficients for both QAP and QKP, and based on these analyses, we examine how small penalty coefficients influence variable fixing.

\subsection{Low-Energy States of the QAP with Small Penalty Coefficients}\label{subsec:discussion-qap-low-energy-states}
In this subsection, we consider the low-energy states of the QAP with small penalty coefficients.
In the small-\( \mu \) limit, \( \mu \to 0 \), the contribution of the constraint term \( \mu \mathcal{H}_{\mathrm{const}} \) becomes negligible, and the system tends to minimize only the objective function \( \mathcal{H}_{\mathrm{obj}} \).
In this regime, several types of assignments can occur.
The trivial all-zero solution \( \bm{x}=\bm{0} \) has \( \mathcal{H}_{\mathrm{obj}}=0 \) and \( \mathcal{H}_{\mathrm{const}}=2L \), which yields \( \mathcal{H}=2L\mu \).
Another possible structure is the single-assignment state, in which only one element \( x_{ik} \) is set to one and the others remain zero.
In this case $\mathcal{H}_{\mathrm{obj}}=0$ and $\mathcal{H}_{\mathrm{const}}=2(L-1)$, so that $\mathcal{H}=2(L-1)\mu$, which is always lower than the all-zero state by $2\mu$.

One can consider partial assignments where \( t (\ll L) \) facilities are assigned to \( t \) distinct locations without conflicts.
In this case the constraint term is \( \mathcal{H}_{\mathrm{const}}=2(L-t) \), leading to \( \mathcal{H}(\bm{x})=\mathcal{H}_{\mathrm{obj}}(\bm{x})+2\mu(L-t) \).
If facility \(b\) is newly assigned to a previously unused location \(a\), the change in the constraint term is \( \Delta \mathcal{H}_{\mathrm{const}}=-2 \), while the increment in the objective function is
\begin{equation}\label{eq:delta-h-obj-qap}
    \Delta \mathcal{H}_{\mathrm{obj}} = \sum_{i=1}^{L} \sum_{k=1}^{L} (d_{ia} f_{kb} + d_{ia} f_{bk}) x_{ik}.
\end{equation}
Thus, the total change is
\begin{equation}
    \Delta \mathcal{H}=\Delta \mathcal{H}_{\mathrm{obj}}-2\mu.
\end{equation}
Only if $\Delta \mathcal{H}_{\mathrm{obj}}<2\mu$ the system can decrease the energy by adding a new assignment; otherwise it remains.

Finally, one can also consider over-assignments, in which some rows or columns contain more than one assignment.
We consider the situation where a new variable \( x_{ab} = 1 \) is introduced in a row that already contains an assignment, that is, \( \sum_i x_{ai} = 2 \).
If \( \sum_j x_{jb} = 1 \) when \( x_{ab} = 1 \), then the change in the constraint term is \( \Delta H_{\mathrm{const}} = 0 \).
On the other hand, if \( \sum_j x_{jb} = 2 \), then the change becomes \( \Delta H_{\mathrm{const}} = 2 \).
The increment of the total energy follows the same form as in \eqref{eq:delta-h-obj-qap}, and hence the energy increment is always non-negative.

From the above classification, it follows that in the limit of small penalty coefficients the dominant structure is the single-assignment state, since it strictly improves upon the all-zero solution, while larger partial assignments require the inequality $\Delta \mathcal{H}_{\mathrm{obj}}<2\mu$ that is rarely satisfied for positive cost matrices, and over-assignments are always inferior to partial ones.
Note that in the partial assignment case, where only \( x_{ik} = 1 \) and all other variables are set to zero, the change in the objective function is given by
\[
    \Delta H_{\mathrm{obj}} = d_{ia} f_{kb}.
\]
If \( i = a \) and \( d_{ii} = 0 \) or \( j = b \) and \( f_{jj} = 0 \), then the energy change for the over-assignment case is \( \Delta H = 0 \).
In contrast, for the partial assignment case, the energy change is
\[
    \Delta H = d_{ia} f_{kb} - 2\mu.
\]
Therefore, when the penalty coefficient \( \mu \) is small, over-assignment may appear preferentially compared to partial assignment.

\subsection{Variable Fixing Behavior of SPVAR and MP-SPVAR in the QAP}\label{subsec:discussion-spvar-mp-spvar-variable-fixing}
In the previous section, we discussed the possible states of the solutions for small values of the penalty coefficient $\mu$ in the QAP.
In this section, we examine how the solutions behave for each $\mu$ and investigate the effect of variable fixing on the solutions.

We consider how the variable fixing behavior of the SPVAR algorithm under different penalty coefficients affects the solution quality.
Two possible reasons can be considered for why the SPVAR method yielded only infeasible solutions when \(\mu\) was small.
First, for small values of \(\mu\), single or partial assignments are likely to occur.
In such cases, some rows or columns may fail to satisfy the one-hot constraint.
Specifically, there may exist rows or columns in which all variables are fixed to zero, thereby leading to infeasibility.
Second, even if no errors occur during the variable-fixing step, the penalty coefficient used in the reduced problem remains small.
Consequently, the solver tends to suppress new assignments beyond those present in the original solution, making it difficult to obtain feasible configurations.
On the other hand, for penalty coefficients large enough to yield feasible solutions, the SPVAR method produced feasible solutions with probability one.
This result suggests that, for such penalty coefficients, neither the approx.\ ratio varies significantly nor does the FS\ ratio deviate from one, indicating that no over-assignments occur and that the variable-fixing step does not violate the two-way one-hot constraint.
In the SPVAR method, variables that take one at least once among all solutions, rather than zero or one in every solution, are retained as variables in the reduced problem.
Since a penalty coefficient that ensures feasibility is also applied in the reduced problem, the final solution can be regarded as the combination of these retained variables that minimizes the objective function.
Therefore, the SPVAR method is expected to provide feasible solutions when the penalty coefficient is sufficiently large.

In the MP-SPVAR method, the variables that can be fixed are considered as follows.
When the penalty coefficient \( \mu \) does not yield any constraint-satisfying solutions, the possible states include single assignments, partial assignments, and over-assignments.
When \( \mu \) allows constraint-satisfying solutions, those solutions themselves are taken as the variables except when all the solutions coincide.
Under the present fixing condition, positions where at least one solution in the solution set takes the value \( 1 \) are treated as variables.
Finally, by resolving the problem using the combinations of the positions treated as variables, the final solution is obtained.
Therefore, when the penalty coefficient \(\mu_{\mathrm{sub}}\) used in the reduced problem is sufficiently large to yield feasible solutions, the MP-SPVAR method can be expected to produce constraint-satisfying solutions.
In this experiment, \(\mu_{\mathrm{sub}}\)  was chosen from the values of \(\mu\) that provided the solutions used for variable fixing.
Therefore, it was crucial that the set included at least one penalty coefficient that yielded feasible solutions.

\subsection{Fixing Behavior of SPVAR and MP-SPVAR in the QKP}\label{subsec:discussion-spvar-mp-spvar-variable-fixing-qkp}
This subsection extends the discussion to the QKP, providing a simplified analysis of variable fixing.
Additionally, the effects on variable fixing behavior by encoding is discussed.
In addition, the effect of the encoding method on the variable-fixing behavior is discussed.
In contrast to the QAP, analyzing the low-energy states of the QKP under small penalty coefficients is more challenging due to the introduction of slack variables, the relative penalty coefficient \( \lambda \), and the specific encoding methods.

When the penalty coefficient is set to zero, the weight constraint is completely ignored, and consequently all items are included.
For small values of the penalty coefficient, the slack variables tend to take values of \( \bm{y} \) such that \( S(\bm{y} ) = 0 \).
As the penalty coefficient increases, the weight constraint is gradually taken into account, while in the optimal solution it is expected that many items will still be included under the weight capacity.
Therefore, the solution should remain close to those with \( S(\bm{y}) = 0 \), which explains why the number of incorrectly fixed slack variables was small for small penalty coefficients.
On the other hand, when the penalty coefficient becomes large, the constraint is enforced more strictly, and the system is likely to select configurations with larger values of \( S(\bm{y}) \).
This leads to an increase in the number of incorrectly fixed slack variables under large penalty coefficients.

Then, we consider the effect of different encoding methods on the slack-variable fixing.
Let the penalty coefficient refer to that associated with the weight constraint, while assuming that the encoding constraints are always satisfied regardless of the penalty coefficient for simplicity.
When the penalty coefficient is small, the integer values represented by the slack variables tend to be small, whereas when the penalty coefficient is large, the integer values of the slack variables are expected to increase.
Under the one-hot encoding, when variable fixing is performed using the SPVAR algorithm across multiple solutions obtained under different penalty coefficients, the variables that take one in any of the solutions are retained as variables with current experimental settings.
With small penalty coefficients, the index of the variable that takes one is smaller than with large penalty coefficients.
Consequently, several weight candidates remain as possible selections for the reduced problem, and the decision variables are determined based on these candidates.
As a result, in the case of one-hot encoding, the proportion of fixed slack variables tends to be large, and the number of incorrectly fixed slack variables is small.
Note that, as discussed in the results, the encoding constraints were not always perfectly satisfied, leading to more complex behavior than the simplified picture above.
Nonetheless, the above discussion provides an intuitive explanation of the observed tendency.

To the contrary, under the domain-wall encoding, the slack variables are represented as a sequence of consecutive ones followed by zeros.
Therefore, under the same assumptions as above, the range of slack variable values retained in the reduced problem spans from the smallest value obtained under the smallest penalty coefficient to the largest value obtained under the largest penalty coefficient.
This is because parts smaller than the minimum slack-variable value are fixed to one, while parts larger than the maximum slack-variable value are fixed to zero.
As a result, the number of variables becomes smaller, whereas the incorrect fix count tends to be higher in the SPVAR algorithm.
Consequently, introducing diversity in the penalty coefficients helps reduce the incorrect fix count in the MP-SPVAR algorithm.
Furthermore, since the fixed ratio of the slack variables is small, it is possible that the variables were not sufficiently fixed, resulting in a limited improvement in the approx.\ ratio compared with the one-hot encoding.

%% file: sections/conclusion/conclusion.tex
In this work, we proposed the multi-penalty sample persistence variable reduction (MP-SPVAR) algorithm as an extension of the SPVAR framework for constrained combinatorial optimization problems (CCOPs).
By exploiting solution information obtained under multiple penalty coefficients, MP-SPVAR demonstrated improved robustness in avoiding incorrect variable fixing compared with the conventional SPVAR method.
Through extensive experiments on benchmark problems, including the QAP and QKP, we showed that MP-SPVAR achieves higher approximation ratios and feasible solution ratios across diverse parameter settings.
Furthermore, the analysis of fixing behavior revealed that incorporating smaller penalty coefficients effectively reduces the likelihood of incorrectly fixed variables, thereby enabling more accurate reduction of the problem size while retaining high-quality solutions.

Despite these promising results, several limitations remain open.
First, while MP-SPVAR improves performance over SPVAR, its performance still depends on the parameter settings.
In particular, the effectiveness is influenced by the selection of penalty coefficient sets and the number of sampled solutions.
Developing systematic strategies for choosing these hyperparameters would further enhance the practicality of the method.
In this study, we fixed the value of the relative penalty coefficient \(\lambda\) within each sample for the QKP.
However, it is also possible to consider applying MP-SPVAR to samples in which diversity in
\(\lambda\) is introduced.
Second, although this study focused on QAP and QKP as representative CCOPs, broader validation on other real-world problems, such as portfolio optimization, scheduling, or logistics, is necessary to assess generalizability.
Third, a fundamental limitation common to all variable reduction methods based on sample persistence, including SPVAR and its extensions, is the loss of persistence in problems with highly degenerate ground states, such as vehicle routing (VRP) and traveling salesman problems (TSP).
In such cases, variables may not consistently exhibit persistent values, leading to little or no variable fixing. Addressing this challenge calls for the development of new variable-fixing algorithms that can operate effectively under degeneracy, for instance by leveraging structural features of solutions, ensemble diversity, or probabilistic persistence metrics.
Finally, the theoretical understanding of the variable-fixing mechanism under diverse penalty regimes remains limited and currently lacks an optimality guarantee.
A more rigorous theoretical analysis could provide deeper insights into why multi-penalty approaches succeed and guide the development of more principled reduction strategies.

Overall, the MP-SPVAR method opens a new direction for enhancing variable reduction in CCOPs. We expect that future work on parameter tuning, application to practical domains, and integration with emerging quantum-inspired hardware as well as quantum algorithms will further advance the potential of this approach toward solving large-scale constrained optimization problems.

%% file: sections/appendicies/appendicies.tex
\begin{figure*}[!t]
    \centering
    \begin{minipage}[ht]{0.24\hsize}\centering
        \includegraphics[width=\linewidth]{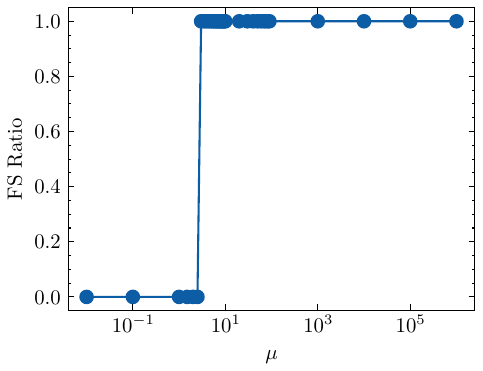}
        \subcaption{}\label{subfig:naive-qap-feasible-tai40b}
        \vspace{0.4em}
        \includegraphics[width=\linewidth]{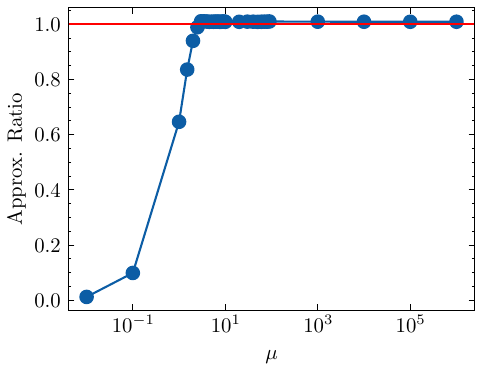}
        \subcaption{}\label{subfig:naive-qap-approx-tai40b}
    \end{minipage}
    \begin{minipage}[ht]{0.24\hsize}\centering
        \includegraphics[width=\linewidth]{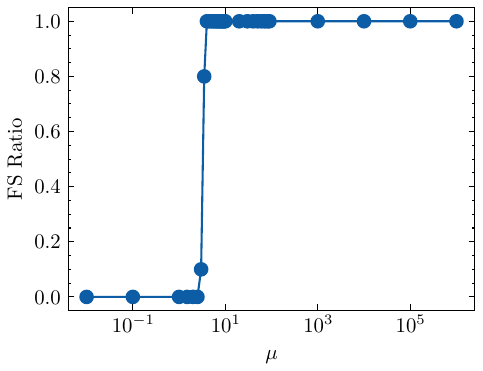}
        \subcaption{}\label{subfig:naive-qap-feasible-tai60b}
        \vspace{0.4em}
        \includegraphics[width=\linewidth]{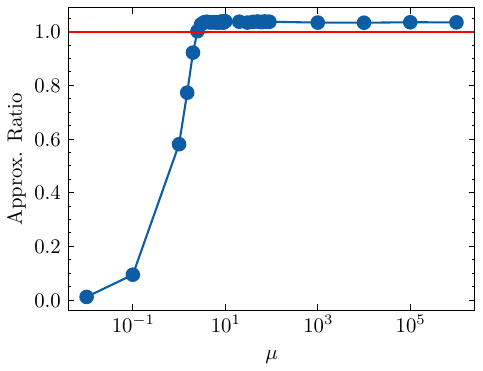}
        \subcaption{}\label{subfig:naive-qap-approx-tai60b}
    \end{minipage}
    \begin{minipage}[ht]{0.24\hsize}\centering
        \includegraphics[width=\linewidth]{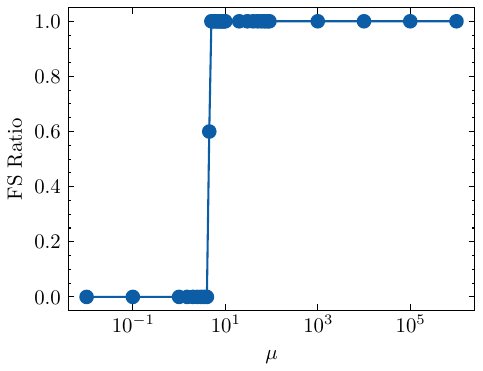}
        \subcaption{}\label{subfig:naive-qap-feasible-tai80b}
        \vspace{0.4em}
        \includegraphics[width=\linewidth]{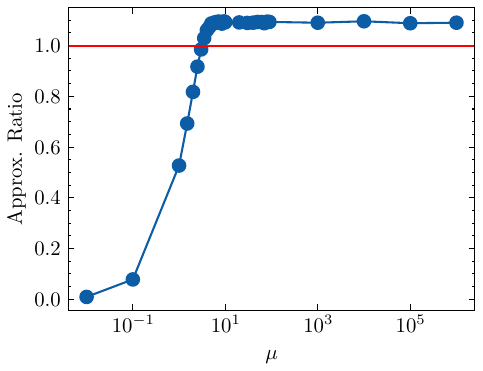}
        \subcaption{}\label{subfig:naive-qap-approx-tai80b}
    \end{minipage}
    \begin{minipage}[ht]{0.24\hsize}\centering
        \includegraphics[width=\linewidth]{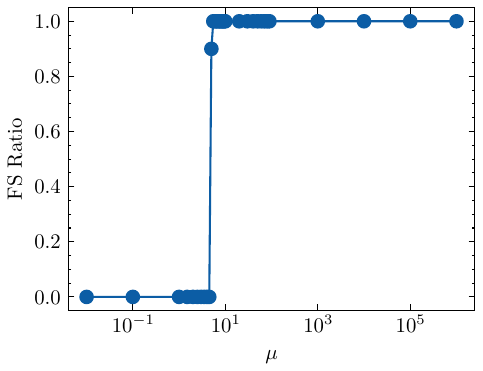}
        \subcaption{}\label{subfig:naive-qap-feasible-tai100b}
        \vspace{0.4em}
        \includegraphics[width=\linewidth]{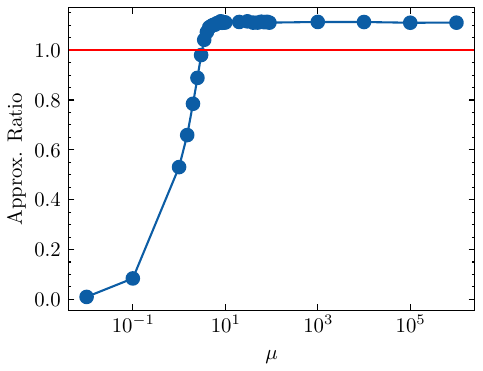}
        \subcaption{}\label{subfig:naive-qap-approx-tai100b}
    \end{minipage}
    \caption{
        (a), (c), (e), and (g) show the FS\ ratio, and (b), (d), (f), and (h) show the approx.\ ratio, as functions of the number of solutions \(N_p\) obtained with the naive method.
        Results are shown for the QAP instances tai40b (40 facilities; (a)--(b)), tai60b (60 facilities; (c)--(d)), tai80b (80 facilities; (e)--(f)), and tai100b (100 facilities; (g)--(h)).
        Each experiment was performed 10 times.
        Here, the approx.\ ratio is also calculated for infeasible solutions.
        The red horizontal line indicates the baseline for approx.\ ratio.
    }\label{fig:naive-qap-tai40b-tai60b-tai80b-tai100b}
\end{figure*}
\section{Penalty coefficient dependency of the naive method on multiple QAP instances}\label{sec:appendix-naive}
In this appendix, we provide experimental results on the penalty coefficient dependency of the naive method on multiple QAP instances.

For QAP, Fig.~\ref{fig:naive-qap-tai40b-tai60b-tai80b-tai100b} shows the FS\ ratio in (a), (c), (e), and (g), and the approx.\ ratio in (b), (d), (f), and (h), as functions of the penalty coefficient obtained with the naive method.
Here, the definition of the approx.\ ratio differs from that in~\eqref{eq:approx-ratio}, as it is also evaluated for infeasible solutions:
\begin{equation}
    \mathrm{Approx.\ Ratio} =
    \frac{\text{Objective Value}}{\text{Best-Known Optimal Value}}.
\end{equation}
The definition of the FS\ ratio remains the same as in~\eqref{eq:fs-ratio}.
As discussed in relation to small penalty coefficients in Section~\ref{sec:discussion}, partial assignments were likely to occur, resulting in objective function values that started from a small baseline.
As the penalty coefficient increased, the approx.\ ratio improved, while the FS\ ratio remained unchanged until it rose sharply between 1 and 10, at which point the probability of obtaining feasible solutions reached one.
Once the penalty coefficient became sufficiently large to guarantee feasibility, the approx.\ ratio ceased to change and stabilized at a constant value.
To verify how good this constant value is, we also evaluated the approx.\ ratio for randomly generated feasible solutions, which exceeded 1.5 across all instances.
The trade-off between the FS\ ratio and the approx.\ ratio with respect to the penalty coefficient is considered to occur within a very narrow range.

\begin{figure}
    \centering
    \begin{minipage}[ht]{0.45\hsize}\centering
        \includegraphics[height=0.115\textheight]{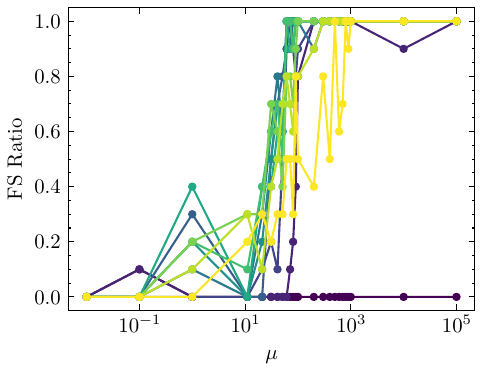}
        \subcaption{}\label{subfig:naive-qkp-jeu_100_25_6-onehot-raw-mu_vs_fsratio}
    \end{minipage}
    \hfill
    \begin{minipage}[ht]{0.54\hsize}\centering
        \includegraphics[height=0.115\textheight]{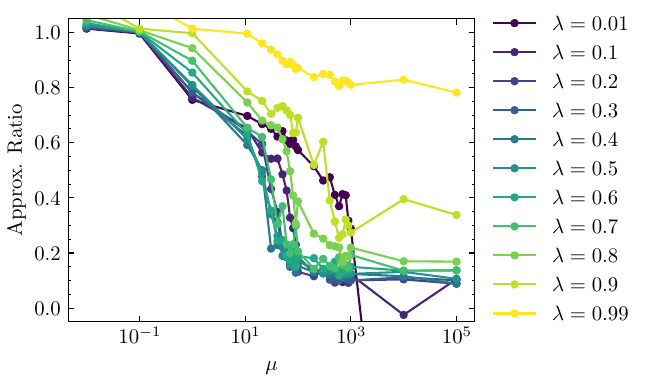}
        \subcaption{}\label{subfig:naive-qkp-jeu_100_25_6-onehot-raw-mu_vs_approx_ratio}
    \end{minipage}
    \begin{minipage}[ht]{0.45\hsize}\centering
        \includegraphics[height=0.115\textheight]{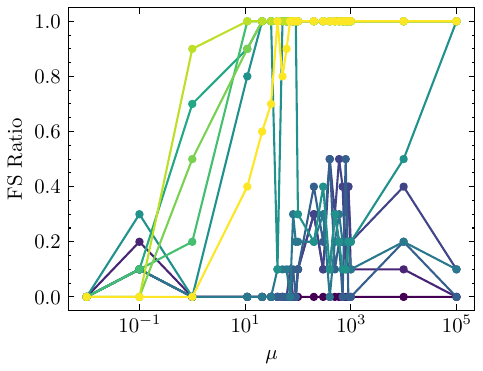}
        \subcaption{}\label{subfig:naive-qkp-jeu_100_25_6-domain_wall-raw-mu_vs_fsratio}
    \end{minipage}
    \hfill
    \begin{minipage}[ht]{0.54\hsize}\centering
        \includegraphics[height=0.115\textheight]{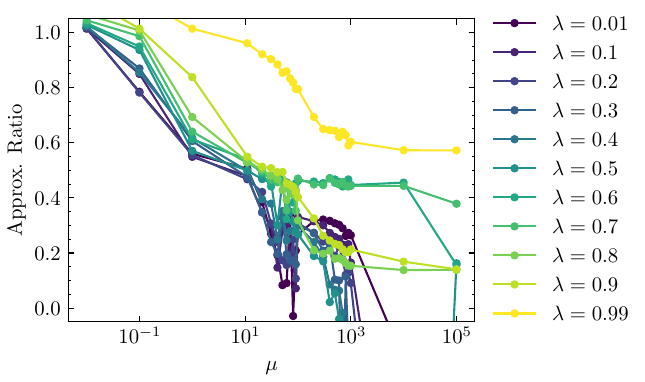}
        \subcaption{}\label{subfig:naive-qkp-jeu_100_25_6-domain_wall-raw-mu_vs_approx_ratio}
    \end{minipage}
    \caption{
        (a), (c) show the FS\ ratio, and (b), (d) show the approx.\ ratio for the 100 items, 1236 capacity QKP instance as functions of the penalty coefficient \(\mu\) obtained with the naive method.
        (a), (b) show the results with the one-hot encoding, and (c), (d) show the results with the domain-wall encoding, respectively.
        The relative penalty coefficient \( \lambda \) was chosen from the set \( \{0.01, 0.1, 0.2, \dots, 0.9, 0.99\} \).
        Each process was performed 10 times.
    }\label{fig:naive-qkp-jeu_100_25_6-mu}
\end{figure}

In the case of the QKP, Fig.~\ref{fig:naive-qkp-jeu_100_25_6-mu} presented the FS\ ratio in (a) and (c), and the approx.\ ratio in (b) and (d), as functions of the penalty coefficient \(\mu\) obtained using the naive method, under a similar experimental setup to that in \cite{Kikuchi2024}.
The experiments were conducted using both the one-hot and domain-wall encodings.
For both encodings, the FS\ ratio increased with \(\mu\), whereas the approx.\ ratio decreased as \(\mu\) increased.
For one-hot encoding, except for \(\lambda = 0.01\) and \(\lambda = 0.99\), the dependence on \(\mu\) showed nearly the same trend, with the FS\ ratio approaching 1.0 for \(\mu \geq 10^{2}\).
For domain-wall encoding, when \(\lambda < 0.5\), the FS\ ratio did not increase even when the penalty coefficient was enlarged.
In contrast, for \(\lambda > 0.6\), the FS\ ratio, similar to that of the one-hot encoding, approached 1.0 for \(\mu \geq 10^{2}\).

\section{Penalty coefficient dependency of the SPVAR and MP-SPVAR algorithms on multiple QAP instances}\label{sec:appendix-spvar}
In this appendix, we provide additional experimental results that complement the analysis presented in Section~\ref{subsec:result-spvar}.

\begin{figure*}[!t]
    \centering
    \begin{minipage}[ht]{0.48\hsize}\centering
        \includegraphics[width=\linewidth]{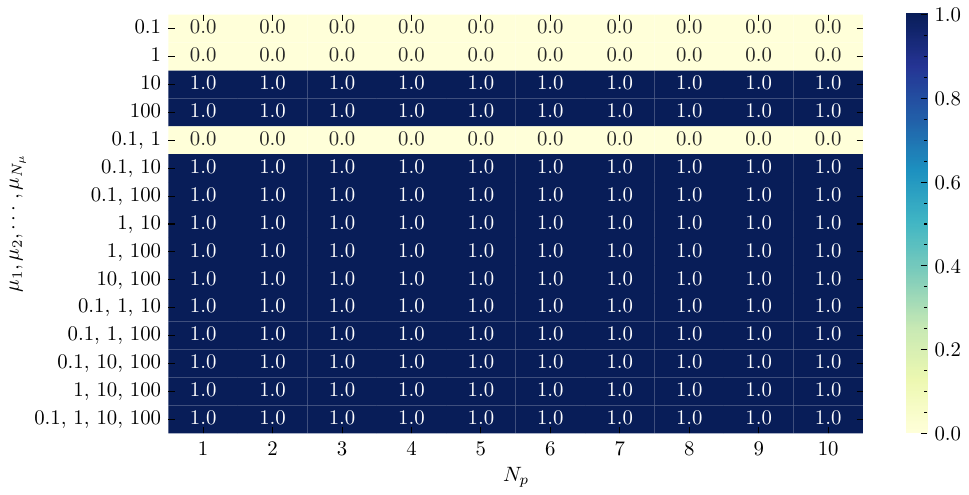}
        \subcaption{}\label{subfig:spvar-qap-feasible-tai40b}
    \end{minipage}
    \hfill
    \begin{minipage}[ht]{0.48\hsize}\centering
        \includegraphics[width=\linewidth]{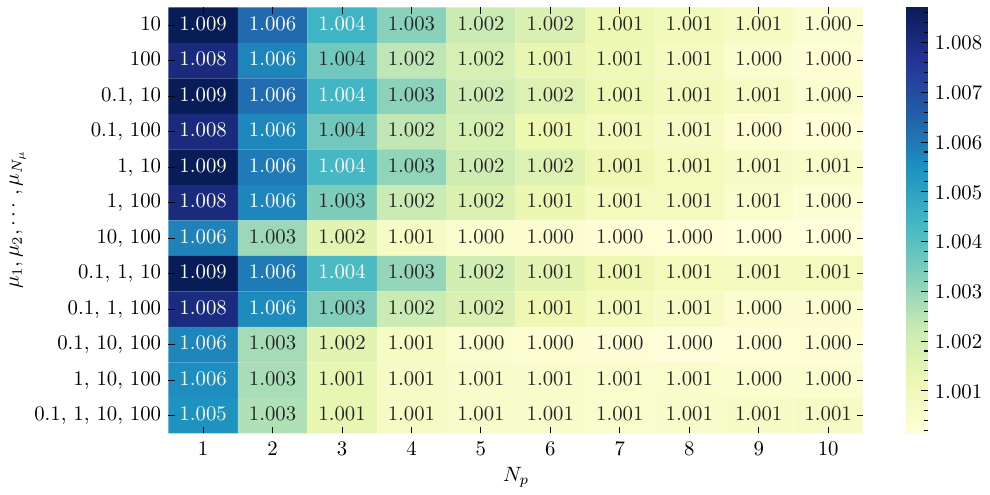}
        \subcaption{}\label{subfig:spvar-qap-approx-tai40b}
    \end{minipage}
    \begin{minipage}[ht]{0.48\hsize}\centering
        \includegraphics[width=\linewidth]{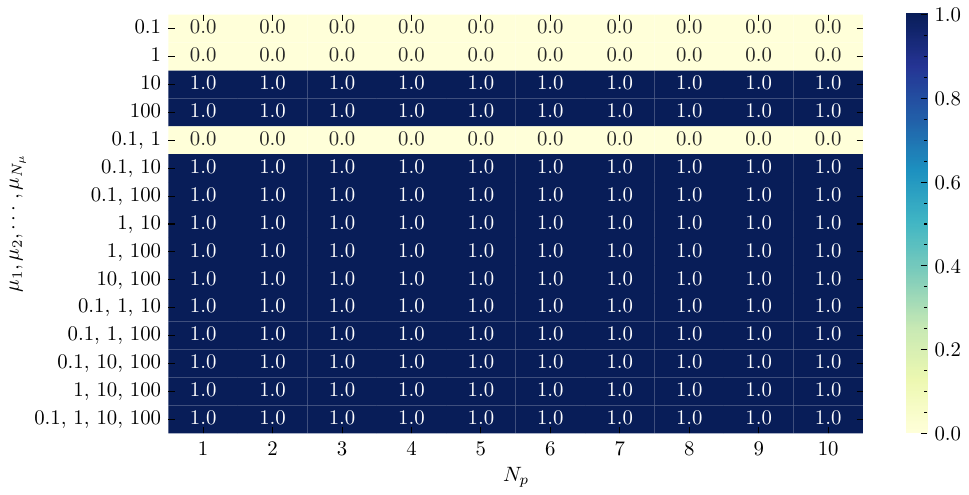}
        \subcaption{}\label{subfig:spvar-qap-feasible-tai80b}
    \end{minipage}
    \hfill
    \begin{minipage}[ht]{0.48\hsize}\centering
        \includegraphics[width=\linewidth]{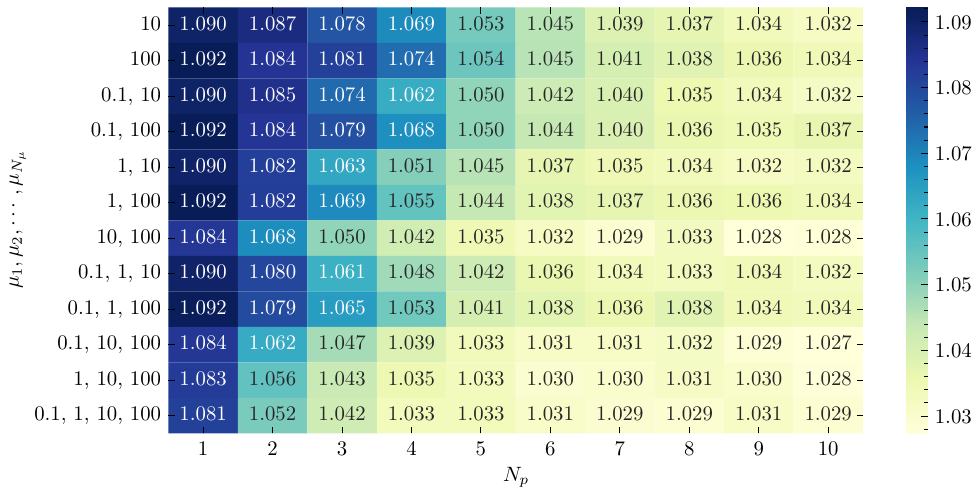}
        \subcaption{}\label{subfig:spvar-qap-approx-tai80b}
    \end{minipage}
    \begin{minipage}[ht]{0.48\hsize}\centering
        \includegraphics[width=\linewidth]{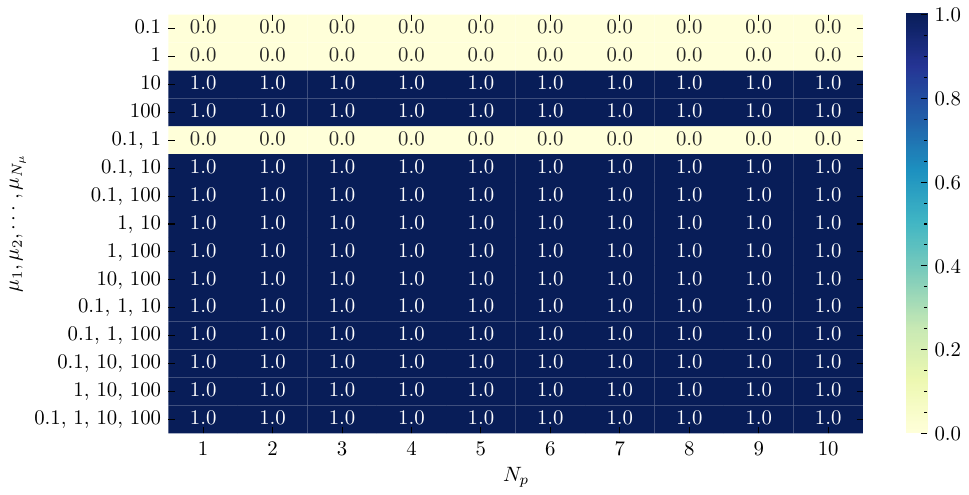}
        \subcaption{}\label{subfig:spvar-qap-feasible-tai100b}
    \end{minipage}
    \hfill
    \begin{minipage}[ht]{0.48\hsize}\centering
        \includegraphics[width=\linewidth]{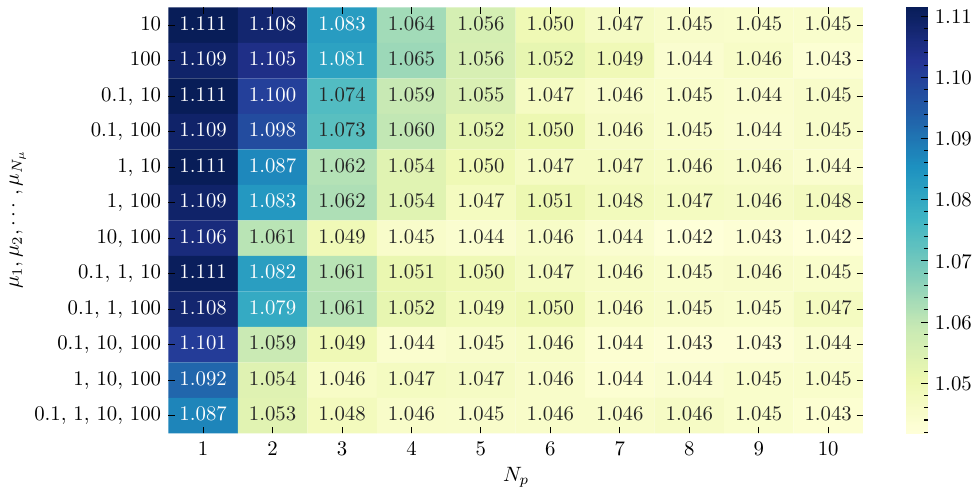}
        \subcaption{}\label{subfig:spvar-qap-approx-tai100b}
    \end{minipage}
    \caption{
        (a), (c), (e) FS\ ratio and (b), (d), (f) approx.\ ratio with respect to the number of solutions \(N_p\) using the SPVAR algorithm.
        (a), (b) show the results for the 40 facility QAP instance called tai40b, (c), (d) show the results for the 80 facility QAP instance called tai80b, and (e), (f) show the results for the 100 facility QAP instance called tai100b.
        Here, \(N_p\) denotes the sample size, and variables that take identical values across all solutions are fixed accordingly.
        The penalty coefficient \(\mu\) is chosen from the set \(\{0.01, 0.1, 1, 10, 100, 1000\}\).
        Each process was performed 10 times.
    }\label{fig:spvar-qap-tai40b-tai80b-tai100b}
\end{figure*}

For the QAP, Fig.~\ref{fig:spvar-qap-tai40b-tai80b-tai100b} presents the approx.\ ratio and FS\ ratio for the benchmark instances tai40b, tai80b, and tai100b, corresponding to 40, 80, and 100 facilities, respectively, obtained from QAPLIB.
These results confirm the general trends discussed in Section~\ref{subsec:result-spvar}.
Specifically, the FS\ ratio was determined solely by the set of penalty coefficients, and in the MP-SPVAR method, feasible solutions were obtained whenever the set included at least one penalty coefficient that yielded feasible solutions in the naive method.
The approx.\ ratio improved as \(N_p\) increased, and when the product of the number of feasible-solution–producing penalty coefficients and \(N_p\) was identical, the MP-SPVAR method achieved an approx.\ ratio comparable to, or slightly better than, that of the SPVAR method.
Furthermore, even small penalty coefficients that did not individually yield feasible solutions contributed to improving the approx.\ ratio.
As the problem size increased, the FS\ ratio remained unchanged, whereas the approx.\ ratio tended to deteriorate.

\section{Relative penalty coefficient \( \lambda \) dependency of the MP-SPVAR algorithm on QKP instances}\label{sec:appendix-mp-spvar}
In this appendix, we show the results of QKP dependency on the relative penalty coefficient \( \lambda \) of the naive method and the MP-SPVAR algorithm.

\begin{figure}
    \centering
    \begin{minipage}[ht]{0.48\hsize}\centering
        \includegraphics[width=\linewidth]{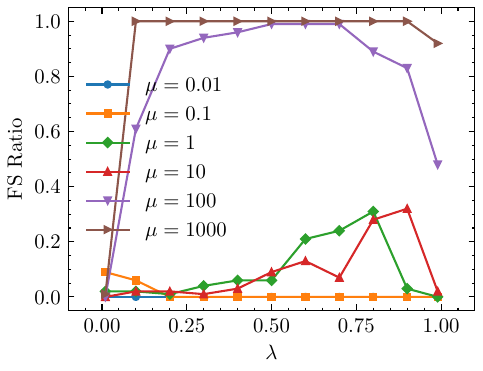}
        \subcaption{}\label{subfig:mp-spvar-qkp-jeu_100_25_6-onehot-raw-fs-ratio}
    \end{minipage}
    \hfill
    \begin{minipage}[ht]{0.48\hsize}\centering
        \includegraphics[width=\linewidth]{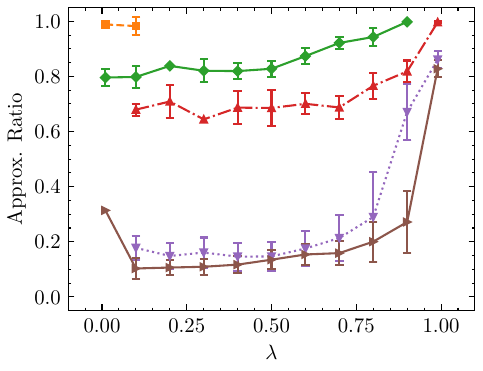}
        \subcaption{}\label{subfig:mp-spvar-qkp-jeu_100_25_6-onehot-raw-approx-ratio}
    \end{minipage}
    \begin{minipage}[ht]{0.48\hsize}\centering
        \includegraphics[width=\linewidth]{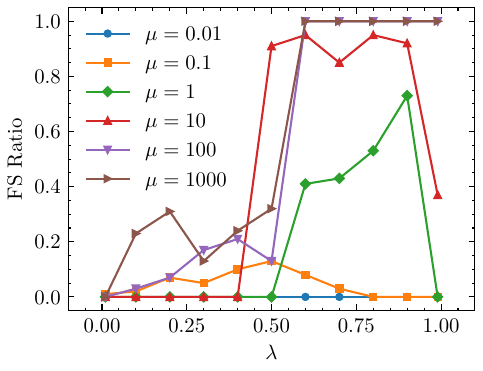}
        \subcaption{}\label{subfig:mp-spvar-qkp-jeu_100_25_6-domain_wall-raw-fs-ratio}
    \end{minipage}
    \hfill
    \begin{minipage}[ht]{0.48\hsize}\centering
        \includegraphics[width=\linewidth]{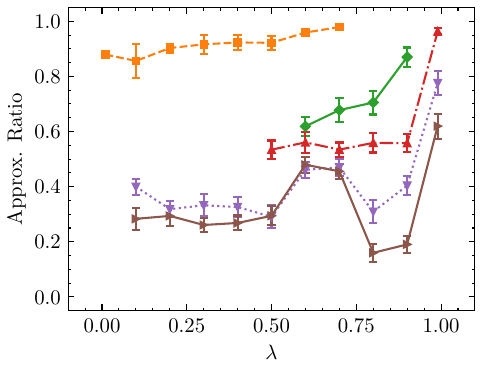}
        \subcaption{}\label{subfig:mp-spvar-qkp-jeu_100_25_6-domain_wall-raw-approx-ratio}
    \end{minipage}
    \caption{
        (a) FS\ ratio and (b) approx.\ ratio with respect to the relative penalty coefficient \(\lambda\) for the 100 items, 1236 capacity QKP instance using the MP-SPVAR algorithm.
        (a), (b) show the results with the one-hot encoding, and (c), (d) show the results with the domain-wall encoding, respectively.
        Each process was performed 10 times.
    }\label{fig:mp-spvar-qkp-jeu_100_25_6-lambda}
\end{figure}

First, in Fig.~\ref{fig:mp-spvar-qkp-jeu_100_25_6-lambda}, we show the results of the approx.\ ratio and FS\ ratio using the naive method with respect to the relative penalty coefficient \( \lambda \) for the 100-item, 1236 capacity QKP instance.
In the case of one-hot encoding, the approx.\ ratio remains nearly unchanged within the range \(0.2 \leq \lambda \leq 0.8\) with large penalty coefficients (e.g. \(\mu = 100, 1000\)).
However, as \(\lambda\) approaches 0 or 1, the approx.\ ratio increases, whereas the FS\ ratio decreases.
In contrast, for domain-wall encoding, the FS\ ratio tends to be higher when \(\lambda > 0.5\) than when \(\lambda \leq 0.5\) with large penalty coefficients (e.g. \(\mu = 100, 1000\)).
From these results, we used \( \lambda = 0.5 \) for one-hot encoding in the MP-SPVAR algorithm because it lies in the stable region and considered as the usual setting.
For domain-wall encoding, it was set to \( \lambda = 0.6 \) because in the case of \( \lambda = 0.5 \) even the large penalty coefficients (e.g. \(\mu = 100, 1000\)) did not yield feasible solutions compared to the case of \( \lambda = 0.6 \).
The trade-off between the FS\ ratio and the approx.\ ratio with respect to the penalty coefficient, as discussed in Appendix~\ref{sec:appendix-naive}, is also observed here.
Moreover, these results indicate the potential for extending the MP-SPVAR method to incorporate diversity in \(\lambda\), as suggested in the future work in Section~\ref{sec:conclusion}.

\begin{figure}
    \centering
    \begin{minipage}[ht]{0.9\hsize}\centering
        \includegraphics[width=\linewidth]{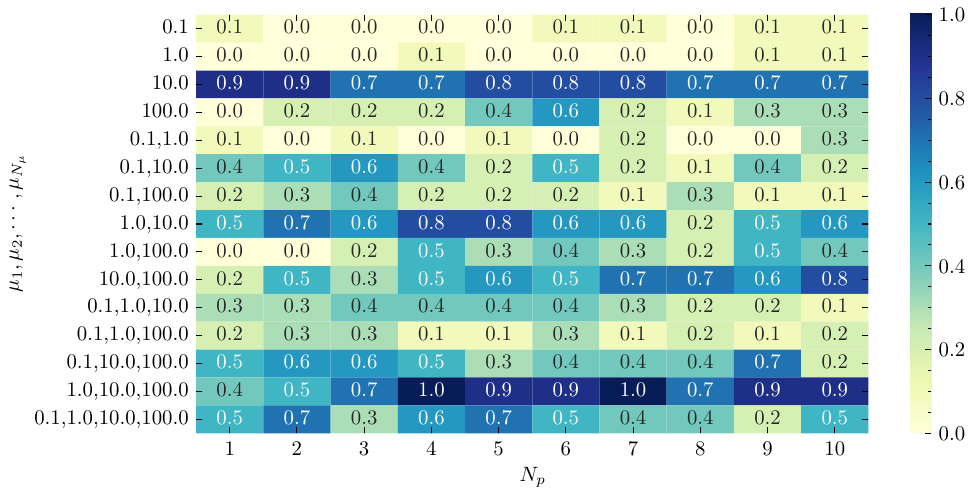}
        \subcaption{}\label{subfig:mp-spvar-qkp-jeu_100_25_6-domain_wall-fs-ratio-lambda_0.5}
    \end{minipage}
    \begin{minipage}[ht]{0.9\hsize}\centering
        \includegraphics[width=\linewidth]{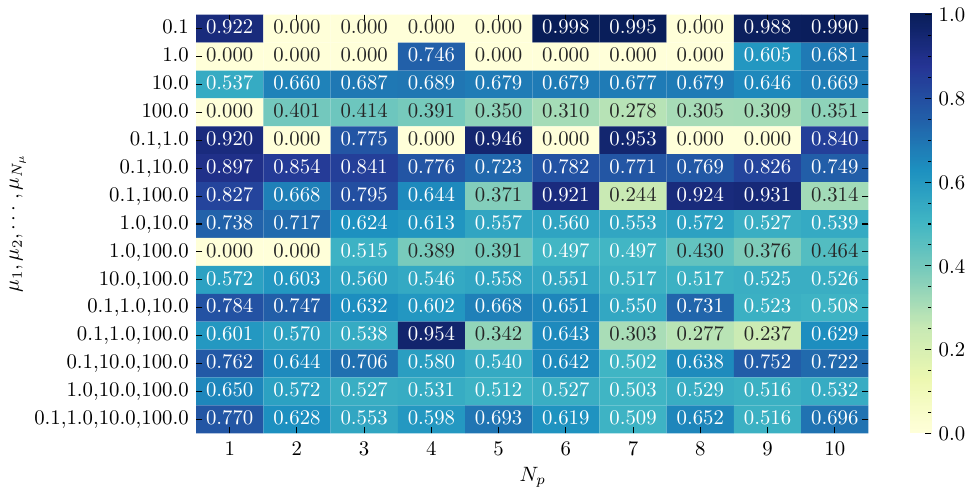}
        \subcaption{}\label{subfig:mp-spvar-qkp-jeu_100_25_6-onehot-approx-ratio-lambda_0.5}
    \end{minipage}
    \caption{
        (a) FS\ ratio and (b) approx.\ ratio with respect to the number of solutions \(N_p\) for the 100 items, 1236 capacity QKP instance using the MP-SPVAR algorithm with the domain-wall encoding.
        Each process was performed 10 times.
        The penalty coefficient \(\lambda\) was set to 0.5.
    }\label{fig:mp-spvar-qkp-jeu_100_25_6-lambda-domain_wall}
\end{figure}

The MP-SPVAR results with domain-wall encoding with \( \lambda = 0.5 \) are shown in Fig.~\ref{fig:mp-spvar-qkp-jeu_100_25_6-lambda-domain_wall}.
From the results of the SPVAR method, the FS\ ratio is higher for \(\mu = 10\) than for \(\mu = 100\).
On the other hand, as discussed in Section~\ref{subsec:result-spvar}, for penalty coefficients that yield feasible solutions with high probability in the naive method ($N_p = 1$), the SPVAR method also produces feasible solutions with similarly high probability.
Moreover, the approx.\ ratio tends to decrease as the penalty coefficient increases.
For the MP-SPVAR method, similar to the discussion in Section~\ref{subsec:result-mp-spvar}, the results simultaneously achieve the approx.\ ratio associated with smaller penalty coefficients and the FS\ ratio associated with larger penalty coefficients.

\section{The Optimal Solution of the QKP instances}\label{sec:appendix-qkp-optimal-solution}
In this appendix, we present the optimal solutions of the QKP instances obtained in our experiments.
As described in Section~\ref{subsec:result-fixing-accuracy}, the benchmark set provides only the objective values of the optimal solutions, but not the explicit solution configurations.
Therefore, we identified the optimal solution for each instance as the configuration obtained during the experiments whose objective value coincided with the known optimal value.
All optimal solutions are available in the online repository~\cite{qkp-opt-dataset}.
The parameter settings used to obtain these solutions are summarized in Table~\ref{tab:qkp-optimal-solution-parameter-settings}.

\begin{table}
    \centering
    \caption{Parameter settings when obtaining the optimal solutions of the QKP instance with the MP-SPVAR algorithm.}\label{tab:qkp-optimal-solution-parameter-settings}
    \begin{tabular}{lc}
        \toprule
        Parameter                                                  & Value         \\
        \midrule
        Instance                                                   & r\_100\_25\_6 \\
        Penalty coefficients \( \mu_1, \cdots, \mu_{N_\mu} \)      & \( 0.1, 1 \)  \\
        Number of solutions for each penalty coefficient \( N_p \) & 4             \\
        Relative penalty coefficient \( \lambda \)                 & 0.5           \\
        The binary-integer encoding method                         & one-hot       \\
        \bottomrule
    \end{tabular}
\end{table}